\documentstyle[aps,epsfig,multicol]{revtex}

\renewcommand{\vec}[1]{{\bf #1}}
\newcommand{\rvec}[1]{\overrightarrow{#1}}
\newcommand{\lvec}[1]{\overleftarrow{#1}}

\newcommand{\be}{\begin{equation}}
\newcommand{\ee}{\end{equation}}
\newcommand{\ber}{\begin{eqnarray}}
\newcommand{\eer}{\end{eqnarray}}

\begin{document}

\def\breakon{\end{multicols}\widetext\vspace{-.2cm}
\noindent\rule{.48\linewidth}{.3mm}\rule{.3mm}{.3cm}\vspace{.0cm}}

\def\breakoff{\vspace{-.2cm}
\noindent
\rule{.52\linewidth}{.0mm}\rule[-.27cm]{.3mm}{.3cm}\rule{.48\linewidth}{.3mm}
\vspace{-.3cm}
\begin{multicols}{2}
\narrowtext}

\draft 

\title{Effective action of a compressible QH state edge: application to 
tunneling.}
\author{L.~S.~Levitov$^{a,c}$, A.~V.~Shytov$^{a,d}$, B.~I.~Halperin$^b$}
\address{{\it (a)} Physics Department, Massachusetts Institute of Technology,
Massachusetts 02139;\\
{\it (b)} Physics Department, Harvard University, Massachusetts 02138;\\
{\it (c)} Center for Materials Science \& Engineering, Massachusetts Institute of Technology,
Massachusetts 02139\\
and Condensed Matter Physics Department, Weizmann Institute of Science, Rehovot 76100, Israel;\\
{\it (d)} L.D.Landau Institute for Theoretical Physics, Moscow 117334, Russia
}

\date{\today}
\maketitle

\begin{abstract}
  The electrodynamical response of the edge of a compressible Quantum 
  Hall system affects tunneling into the edge. Using the composite 
  Fermi liquid theory, we derive an effective action for the edge modes 
  interacting with tunneling charge. This action generalizes the chiral
  Luttinger liquid theory of the Quantum Hall edge to compressible systems
  in which transport is characterized by a finite Hall angle. 
  In addition to the standard terms, the action
  contains a dissipative term. The tunneling exponent is calculated as
  a function of the filling fraction for several models, including 
screened and unscreened long-range Coulomb interaction, 
as well as a short-range interaction. 
  We find that tunneling exponents are robust and to a large extent 
  insensitive to the particular model. 
  We discuss recent tunneling measurements in the overgrown cleaved edge 
  systems, and demonstrate that the profile of charge density near the edge 
  is very sensitive to the parameters of the system. In general, the density
  is nonmonotonic, and can deviate from the bulk value by up to $30\%$. 
  Implications for the correspondence to the chiral Luttinger edge theories 
  are discussed.

\end{abstract}
\bigskip

\begin{multicols}{2}
\narrowtext
%
%
%
\tableofcontents

\section{Introduction}
\subsection{Background and recent work}

The edge of a Quantum Hall (QH) system attracts a lot of interest
because it provides an example of a one dimensional non-Fermi-liquid.
The theoretical picture of QH edge was first developed for odd-denominator 
Landau-level
filling fractions $\nu$ that correspond to incompressible QH states\cite{Wen}. 
It involves one or several interacting chiral Luttinger liquid modes. 
The most prominent feature of Luttinger liquid is the power law character of 
the Green's function.

A power like Green's function leads to a power law in
the tunneling-current--voltage dependence: $I  \sim V^{\alpha}$. 
The tunneling exponent $\alpha$ has been extensively 
studied theoretically for the principal filling fractions of Laughlin 
and Jain hierarchies\cite{Wen,Fisher}. For Laughlin states with 
$\nu=1/(2k+1)$ the edge is described by one chiral mode and tunneling 
current $I \sim V^{2k +1}$ is predicted\cite{Wen}. Theories of the edge 
with $\nu \neq 1 / (2k+1)$ involve more than one  mode. In the multi-mode 
case the results are qualitatively different for the modes
going all in one direction, and for modes going in the opposite 
directions.

For comoving edge modes, the tunneling exponent is universal and does
not depend on the character of interaction between the modes. For
example, this is the case at the Jain filling fractions $\nu=n/(n
p+1)$ with positive integer $n$ and even $p$, where Wen\cite{Wen} finds
$I\sim V^{p+1}$. On the other hand, for the edge described by
modes going in opposite directions, the tunneling exponent
depends on the interaction strength. In this case, it is also
important to take into account the effects of disorder\cite{Fisher}. The
point is that relaxation between the modes due to scattering by
disorder mixes the modes, and at sufficiently high disorder
effectively forms a single charged mode with universal tunneling
exponent. For example, for the Jain fractions $n/(np+1)$ with
$n<0$ and even $p>0$, Kane, Fisher, and Polchinski\cite{Fisher} 
found $I\sim V^{p+1-2/|n|}$.

Tunneling experiments probing the physics of the QH edges were 
first attempted using conventional split gate devices\cite{Webb},
after which a new generation of 2D systems was developed by using the
cleaved  edge  overgrowth technique\cite{Pfeiffer'90}. In these
structures  it is possible to study tunneling into the edge of a
2D electron gas from a 3D doped region. In this system  one  can
create a 2DEG with a very sharp edge, with residual 
roughness of an atomic scale. High quality of the cleaved edge system
permits to explore tunneling in both incompressible and compressible QH
states\cite{Chang2,Grayson}. 
		   
In the first experiment\cite{Chang2}, for $\nu=1/3$ it was found that
the tunneling conductivity is non-ohmic, $I\sim V^\alpha$, 
with the exponent $\alpha\simeq 2.7$, quite close to 
the theoretical prediction $\alpha=3$. After that, it was 
observed\cite{Grayson} that the power law
$I\sim V^\alpha$ holds for both incompressible and compressible QH densities, 
in the range $0.25<\nu<1$. The exponent $\alpha$ was found to be reasonably 
accurately given by a simple formula: $\alpha=1/\nu$. 
Interestingly, this dependence did not
agree with the predictions of chiral Luttinger models, except
for a special point $\nu=1/3$. Moreover, it was quite surprising
that the power law is equally well obeyed by both compressible and 
incompressible values of $\nu$. 

These findings prompted interest in the problem of
tunneling into the edge of a compressible QH system. 
A good description of the compressible QH states is provided by 
the composite fermion theory. This theory\cite{HLR} is constructed at 
$\nu=1/2$
and other rational $\nu$ with even denominator, 
and is used to map the problem of fractional QH effect onto
the integer QH problem for new quasiparticles, composite
fermions interacting via statistical Chern-Simons gauge field.
In the composite fermion picture, an electron is described as a
fermion carrying vorticity represented by a quantized gauge
field vortex. For densities close to the half-filled Landau level the vortex 
has
$p=2$ flux quanta. The theory of composite fermions with $p=2$ 
describes the interval of densities $1/3<\nu<1$. At smaller densities 
$1/5<\nu<1/3$
composite fermions with $p=4$ are used, etc.

The theory of tunneling into a compressible QH edge\cite{SLH} which uses
composite fermions to describe the QH system predicts the power 
law $I\sim V^\alpha$, with $\alpha$ being a continuous function of $\nu$:
  \be\label{Alpha_limit}
 \alpha = 1
        + {e^2\over h}
          \Big(
               \rho_{xy} - |\rho^{(0)}_{xy}|
          \Big)
         , 
 \ee
where $\rho_{xy}=(h/e^2)\nu^{-1}$ is the Hall resistance 
of the 2DEG and $\rho^{(0)}_{xy}=(h/e^2)(\nu^{-1}-p)$ is the Hall
resistance of composite fermions moving in an effective magnetic field $B_{\rm 
eff}=B-p\Phi_0 n$,
where $n$ is electron density. The result (\ref{Alpha_limit}) describes the 
system
in the limit $\rho_{xx}\to 0$.
The dependence of $\alpha$ on $1/\nu$ is monotonic, and is characterized 
by plateaus in the intervals $2<1/\nu<3$, $4<1/\nu<5$, etc. (see 
Fig.~\ref{fig1} below).
The plateaus are connected by straight lines with slope $2$. The cusp-like 
singularities
predicted in the dependence $\alpha(\nu)$ at $\nu=1/2,\,1/4$, etc., are 
somewhat
smeared when $\rho_{xx}$ is finite (see Eq.~(\ref{Alpha}) and 
Fig.~\ref{fig1}).

Interestingly, these results match exactly the predictions
of the chiral Luttinger liquid theory for Jain series of
incompressible states. Although formally this theory lacks continuity in the 
filling
fraction, starting from a new set of edge modes for each given filling 
fraction,
the exponents $\alpha(\nu)$ obtained by Wen for $\nu=n/(pn+1)$ and by 
by Kane, Fisher and Polchinski for $\nu=n/(pn-1)$ fall on the continuous 
curve (\ref{Alpha_limit}) found in the composite fermion calculation.
The exponents of Wen fall on the plateaus, while
those of Kane et al. fall on the straight lines connecting 
the plateaus.

However, the disagreement with the experimentally measured $\alpha(\nu)$
requires an explanation. Recently a number of theories were 
proposed trying to resolve this issue. 
In one approach, described by Conti and Vignale\cite{ContiVignale}, 
Han and Thouless\cite{HanThouless} and Z\"ulicke and 
MacDonald\cite{ZulickeMacDonald},
tunneling is studied by using a hydrodynamical theory of a compressible QH 
edge, in which
the nature of underlying quasiparticles is essentially ignored. From such a
treatment the desired relation $\alpha=1/\nu$ emerges readily, as we will
discuss in detail below in Sec.~\ref{sec-phase} and 
at the end of Sec.~\ref{sec-simple}. However, this approach 
ignores the contribution to the electron Green's function of the quasiparticles 
in the QH state,
and thus it is in contradiction with 
the presently existing microscopic picture of the QH effect. 

Another line of thought, developed 
by Alekseev {\it et al.}\cite{Cheianov}, is that the experimental system is 
not what
it is assumed to be. In particular, it is proposed that instead 
of a clean edge the real system contains many localized states
in sufficient proximity to the edge. Then, if one assumes that
the tunneling rate bottleneck corresponds to tunneling from the doped region 
into
a localized state, and that the density of localized states is sharply peaked 
about
the Fermi energy, one finds the desired result $\alpha = 1/\nu$.
The reason is that in the problem involving a localized state
no conversion of an electron into a quasiparticle 
is required, and the only effect to be considered is a shakeup of the
edge plasmon mode, the effect equivalent to the x-ray edge problem 
in the Fermi liquid.  However, it is not clear why the density of localized 
states
should be peaked at the Fermi energy in the actual samples.
An apparently similar idea has been developed earlier 
by Pruisken {\it et al.}\cite{Pruisken} using quite elaborate methods which
we have not been able to follow in detail.

Also, a theory using composite fermions was proposed by D.-H. Lee and X.-G. 
Wen\cite{LeeWen}
in which both charged (edge plasmon) and neutral (quasiparticle) modes
are included. It was assumed, however, that the velocity of the charged mode 
is much larger
than the neutral mode velocity. In this case, there exists an intermediate 
energy regime in which
only the charge mode dynamics is important, while the neutral mode response 
can be ignored.
In this energy domain one obtains $\alpha=1/\nu$. It should be pointed out, 
however, that
the ratio of the charged and neutral mode velocities is of order of $\ln 
w/r_s$, where $w$
is the distance from the edge to the doped region, and $r_s$ is the screening 
radius.
Optimistically, the ratio $w/r_s$ can be as large as $10$ which is still not 
enough to explain the
power law demonstrated in a wide range of $2.5$ decades in the bias voltage. 

Another approach trying to rationalize the measured tunneling exponent
$\alpha\sim 1/\nu$ was proposed by Khveshchenko\cite{Khveshchenko}. 
This theory is based on composite fermions and is similar 
in its assumptions 
to Ref.\cite{SLH} and to the present work. However, the calculated 
tunneling exponent is $1/\nu$ up to a frequency dependent logarithmic correction
small in $\rho_{xx}/\rho_{xy}$. We believe that this is due to an inconsistency
of the analysis ignoring important effects accounting for dynamics of free
composite fermions. One can see that by comparing
Eq.7 of Ref.\cite{Khveshchenko} with our Eq.\ref{phase-SLH-answer}, and noting
that the term describing the free composite fermion response is missing in 
Ref.\cite{Khveshchenko}.

In addition to this controversy, the theory by Lopez and Fradkin\cite{LopezFradkin} seems to abandon
the entire theoretical picture of the multi-mode QH edges proposed in
Refs.\cite{Wen,Fisher} for the incompressible Jain fractions. 
Unlike Ref.\cite{LeeWen}, the
authors of Ref.\cite{LopezFradkin} are not using a microscopic mechanism for eliminating
the neutral propagating edge modes. The construction proposed in Ref.\cite{LopezFradkin} 
involves only one charged mode plus two auxiliary Klein factors which do not constitute
additional propagating degrees of freedom. In that, the approach of Ref.\cite{LopezFradkin}
can be comapred to the conventional quantum Hall edge theories\cite{Wen,Fisher}
in which the velocity of neutral mode is exactly zero. If true, this would lead to 
the $\alpha=1/\nu$ dependence at arbitrarily low energies. However, it is presently
unclear whether the picture of the neutral mode with zero velocity 
can be justified microscopically. 

What complicates the controversy even further is the recently presented 
evidence
of a plateau-like feature exhibited by $\alpha(\nu)$ in some 
cleaved edge samples\cite{Chang-plateau}. The value of $\nu$ near which
the dependence $\alpha(\nu)$ flattens out is however quite close to $1/3$, 
whereas
the expected plateau interval is $2<\nu^{-1}<3$. This discrepancy may be 
explained
by solution of the electrostatic problem near the edge 
(see Sec.~\ref{sec-experiment} below and Ref.~\cite{Chang-plateau}) 
which shows that in a wide region adjacent to the edge the density exceeds 
the bulk value by about $20-30\%$. 
Because of this behavior of the density profile,
the feature in $\alpha(\nu)$ observed near $\nu_{\rm bulk}=1/3$ may 
correspond to somewhat higher density near the edge, with
$\nu$ somewhere between $1/3$ and $1/2$. 

One other complication is that the analysis of the electrostatic problem shows 
that
the density profile near the edge can be nonmonotonic and, in general, 
depends quite sensitively on the system parameters. This observation can make 
the relation with the theories assuming constant filling factor somewhat 
indirect.
At present, the matter is obviously far from being resolved, and more 
experimental
and theoretical studies are needed to clarify the situation. With this in 
mind,
in this article we present an alternative derivation of the results obtained 
in Ref.\cite{SLH}, demonstrating their robustness and establishing a more 
direct
connection with the chiral Luttinger theories of the QH edge. 

The basis of our analysis will be the theory of composite fermions\cite{HLR}. 
We assume that noninteracting composite fermions are characterized
by $\rho^{(0)}_{xx}$ and $\rho^{(0)}_{xy}$ which may depend,
e.g., on the filling fraction. The measured resistivities are
then $\rho_{xy}= \rho^{(0)}_{xy}+p h/e^2$ and $\rho_{xx}=
\rho^{(0)}_{xx}$, where $p$ is the number of flux quanta
attached to an electron ($p=2$ for $1/3 < \nu < 1$). 
The tunneling current is expressed in a standard way in terms of the
electron Green's function. We derive the relation between Green's functions
of an electron and of a composite fermion, and compute the former 
using a ``factorization approximation.'' In this analysis the effects
of shaking up slow electromagnetic and Chern-Simons gauge field modes are
separated out. As a result, the tunneling current is expressed 
in terms of electromagnetic response functions and the number of flux quanta 
$p$.
The theory predicts a power law $I\sim V^\alpha$ with a continuous 
dependence of the tunneling exponent $\alpha$ on the filling fraction. 
As far as tunneling into the edge is concerned, 
there is no qualitative difference between 
compressible and incompressible states. The ``Luttinger liquid-like''
behavior in the edge tunneling emerges when the Hall angle is close to $\pi/2$,
for both compressible and incompressible electron systems. 

The paper is organized in the following way. 
In Sec.~\ref{sec-phase} we review the approach of Ref.\cite{SLH} 
based on a semiclassical phase factor analysis of the Green's function.
This is done with the purpose of motivating and providing connection 
with the subsequent discussion of the effective action formalism. 
In Sec.~\ref{sec-model} we 
begin laying out the basic approach of the present theory of tunneling. 
At low energy, the most important effect is the shake-up of long wavelength 
modes corresponding to spreading of the tunneling charge. 
To describe it, one can use a semiclassical method, which provides 
a simple and universal picture of tunneling\cite{LevitovShytov}. 
We then construct an effective action
in $D=2$ written in terms of composite fermion density and current, 
as well as the Chern-Simons gauge field. Section~\ref{sec-model} ends by
proving an important identity for this action which is used in 
the following part of the paper. 

In this paper we focus on the relatively simple 
``dirty composite fermion'' case, 
corresponding to
composite fermions scattered by the disorder, and described by 
finite Ohmic conductivity.
In Sec.~\ref{sec-simple} we consider the problem with short-range interaction
between composite fermions. In the $D=2$ action we integrate 
over the variables in the bulk and derive an effective $D=1$ action 
that describes the dynamics in terms of the variables at the edge. 
This action is basically of chiral Luttinger form, with an extra 
``dissipative'' term
nonlocal in space and time, which takes into account the effects of charge 
relaxation in the bulk.
The $D=2\ \to\ D=1$ reduction for the problem with 
short-range CF interaction can be handled in an elementary way
and leads to a simple algebraic expression for the tunneling exponent
in terms of Ohmic and Hall conductivities. 

Then, in Sec.~\ref{sec-coulomb} we repeat the analysis for the problem 
with long-range Coulomb interaction.
In this case the $D=2\ \to\ D=1$ reduction procedure involves solving a 
boundary value problem
for dynamical screening near the edge. We consider three different models, 
describing the problems with
unscreened Coulomb interaction and also taking into account screening
due to image charges induced in the doped overgrown region. 
(This screening has the peculiarity that the screened interaction remains 
long ranged,
because the image charges are located not above the 2DEG, but on the side of 
the 2DEG edge.)
Two of these boundary value problems can be solved by
elementary methods using Fourier transform, and one leads to
an integral equation of Wiener-Hopf type. 
In all three cases, we use the effective $D=1$ action to compute the 
tunneling current, and derive an expression for the tunneling exponent 
$\alpha$.

In the case of unscreened interaction the tunneling exponent $\alpha$ 
turns out to be somewhat frequency 
dependent, having a contribution proportional to $\rho_{xx}\ln\omega$, which 
corresponds to a slight deviation from the power law. 
However, in the most realistic of the three models
accounting for screening by the doped region, we find a nearly perfect power 
law.
Otherwise, the results for the three models with long-range interaction, 
screened and unscreened, and for the short-range interaction model, give 
essentially
the same dependence of the tunneling exponent on $\rho_{xy}$, and thus all 
agree.
The agreement of the results for different kinds of interaction implies that
they are robust. 

In the calculations described above, we characterize the system by
a resistivity tensor that is
independent of wave vector and frequency.  In particular, this assumption
implies that we are
restricted to tunneling at voltages and temperatures small compared to the
scattering rate of the composite fermions. At energies above the scattering
frequency, but below the Fermi energy, one is in a different regime  (the
``clean regime'') where  ballistic dynamics should be used for the
composite fermions.  This regime may be of considerable practical interest 
because
the samples used for the tunneling measurements are of very high mobility,
and are presumably quite clean even near the edge.
Even for electron energies below the CF scattering frequency, however, one
should really check that contributions from wave vectors larger than the inverse
mean free path can safely be ignored.   

A proper treatment of the ballistic region requires the use of nonlocal
electromagnetic response functions, and is considerably more difficult than
the models discussed in the present paper. In the Appendix~\ref{ballistic} below 
we
investigate a simplified model for the nonlocal conductivity, which
serves to illustrate some of the salient features of the problem.  The
simplified model is not adequate, however, to answer unambiguosly the
fundamental theoretical question: whether low-energy degrees of freedom at
short length scales can significantly alter the tunneling exponent at low
electron energies.

In order to better address this problem, we have also undertaken a 
numerical solution of the charge spreading problem with a proper
representation of the nonlocal conductivity. 
Preliminary results suggest that the tunneling
exponents will not be changed by a large amount from the results obtained
in the present paper\cite{ShytovThesis}, but further work is necessary here.

One should also recall that in the limit of very low temperatures and
frequencies, in compressible states, one expects that there will be
interaction corrections to the resistivity itself which depend
logarithmically on energy \cite{HLR}.  Therefore, in principle, 
at sufficiently low energies, the
renormalized value of $\rho_{xx}$ will become comparable to the value of
$\rho_{xy}$ and our entire analysis may cease to be valid.  However, the
energy range where this would occur is too small to be of experimental interest
in high-mobility samples where the bare value of
$\rho_{xx}$ is small.

%
%

\subsection{The semiclassical phase method}
\label{sec-phase}

The tunneling exponent (\ref{Alpha_limit}) was derived in Ref.\cite{SLH}
using a ``semiclassical phase'' approach. 
Here we restate the derivation of Eq.(\ref{Alpha_limit}) emphasizing
the connection with the effective action method being used 
in the main part of this article.

One advantage of the semiclassical phase method employed 
in Ref.\cite{SLH} is that 
it does not require subtraction of counterterms like 
${\cal S}-{\cal S}_{\rm free}$
used in the following sections. 
A suspicious reader may think of this subtraction as  
an {\it ad hoc} procedure motivated only on physical grounds. 
Although we justify the counterterms subtraction carefully and rigorously
below in Sec.~\ref{sec-identity}, it will perhaps be useful for the reader 
to
see the same result derived by an alternative method. 

It should be mentioned that the phase method, although more appealing 
intuitively,
is more difficult in use, especially in problems with the boundary, like 
the edge tunneling problem. Because of that our use of it here 
is limited to the simplest case when the interaction is solely due 
to the Chern-Simons gauge field, and there is no long-range Coulomb 
interaction.
The short-range interaction is assumed to be taken into account by the 
composite fermion transformation.

We start with the tunneling electron Green's function in imaginary time. 
One can formally write it as an average over the fluctuations of the gauge 
field:
  \breakon
  \be\label{phase-G-general}
G_{{\bf r}\,{\bf r'}}(t_{12}) =
{\cal Z}^{-1} \int {\cal D}\{a_{\mu}\} \, G_{{\bf r} \, {\bf r'}}
(t_1, t_2, \, a_{\mu}) \,
  e^{-{\cal L}_{eff}[a_\mu]}
 ,\qquad
t_{12}=t_2-t_1
 .
  \ee 
  \breakoff\noindent
This exact expression emphasizes the order of integration over fermionic 
fields and
the gauge field $a_{\mu}$. Here $G_{{\bf r} \, {\bf r'}}(t_1, t_2, \, a_{\mu})$
is the electron Green's function for a given configuration of the gauge field 
$a_{\mu}(\vec r,t)$.
For evaluating the tunneling current, we will need $G_{{\bf r}\,{\bf 
r'}}(t_{12})$
for ${\bf r} = {\bf r'}$.

The effective action ${\cal L}_{eff}[a_\mu]$ is the RPA action derived 
in Ref.\cite{HLR}. Below we will only need ${\cal L}_{eff}$ up to 
quadratic order:
  \be\label{phase-L-eff}
{\cal L}_{eff}[a_\mu]=\frac{1}{2}\int a_\mu(x){\cal 
D}^{-1}_{\mu\nu}(x,x')a_\nu(x')\, d^3x d^3x'
  \ee
where the correlator of gauge field fluctuations
${\cal D}_{\mu\nu}(x,\, x') =
\langle a_{\mu}(x)\, a_{\nu}(x') \rangle$
for the CF system in the absence of long-range interaction 
in the RPA approximation\cite{HLR} is given by 
  \be
{\cal D}^{-1}_{\mu\nu}(k)=
{\cal K}_{\mu\nu}(k)+\frac{i}{4\pi p}k^\lambda\epsilon_{\mu\nu\lambda}
  \ee
Here ${\cal K}_{\mu\nu}=\langle j_{\mu} j_{\nu} \rangle$ is the free fermion 
current correlator
(cf. Ref.\cite{HLR} and Sec..~\ref{sec-model} below).

We employ a semiclassical approximation for
$G_{\bf r\,r}(t_1, t_2,\,a_{\mu})$. To motivate it, think of an injected
electron which rapidly binds $p$ flux quanta and turns into a
composite fermion. The latter moves in the gauge field $a_{\mu}$
and picks up the phase
  \be
\phi[a_{\mu}] = \int^\infty_{-\infty} a^{\mu}({\bf r},\, t)\, 
                         j^{\rm free}_{\mu}({\bf r}, \, t)\,
                    d^2{\bf r}\, dt ,
  \ee
where $j^{\rm free}_{\mu}({\bf r}, \, t)$ is the current describing
spreading of {\it free} composite-fermion density.
Semiclassically in $a_{\mu}({\bf r},\, t)$, one writes
  \be\label{phase-G}
G_{{\bf r}\, {\bf r}}(t_1,t_2, \, a_{\mu}) =
    e^{i\phi[a_{\mu}]}\, G^{(0)}(t_{12})
 ,
  \ee
where $G^{(0)}(t_{12})\simeq t_{12}^{-1}$ 
is the composite-fermion Green's function in
the absence of the slow gauge field. Note that fast fluctuations of
$a_\mu$ are included in $G^{(0)}(t)$ through renormalization of
Fermi-liquid parameters.  

Let us remind the reader that electron Green's function in the composite 
fermion theory has an additional phase factor 
$\exp\left(i\int^{t_2}_{t_1}a_0(t')dt'\right)$ 
introduced by Kim and Wen\cite{KimWen} 
which accounts for the gauge field of a solenoid inserted into 
the system upon the transformation
of the tunneling electron into a composite fermion. This phase factor
has been discussed in the context of the problem of 
tunneling into the
bulk. By virtue of gauge invariance of electron Green's function under gauge
transformations of the Chern-Simons field, one can eliminate the phase factor 
using the Weil gauge $a_0=0$. Because of that, seemingly different 
approaches to the bulk tunneling problem, some emphasizing the phase 
factor\cite{KimWen} and others ignoring it\cite{HePlatzmanHalperin,LevitovShytov}, 
are essentially equivalent. Below we are going to use the $a_0=0$ gauge, 
which permits us to drop the solenoid phase factor from the start.

Now, we substitute the Green's function in the phase approximation 
(\ref{phase-G})
into Eq.(\ref{phase-G-general}) and average over fluctuations of $a_\mu$ using
the action (\ref{phase-L-eff}). This gives
  \be
G_{{\bf r}\, {\bf r}}(t)=G^{(0)}(t)e^{-S}
  \ee
where
  \be
S = \frac{1}{2}
\int d^3x\, d^3x'\, j_{\mu}^{\rm free}(x)\,
j_{\nu}^{\rm free}(x')\,\,
{\cal D}_{\mu\nu}(x, x')  .
  \ee
It is convenient to rewrite the exponent $S$ hereafter called ``action'' 
as follows
  \be\label{phase-ja-RPA}
S = -\frac{i}{2}
\int d^3x j_{\mu}^{\rm free}(x)\tilde a^\mu(x)
  \ee
where $\tilde a_\mu(x)=i\int {\cal D}_{\mu\nu}(x, x') j_{\mu}^{\rm free}(x') 
d^3x'$
is the actual gauge field produced by the moving charge. The representation 
(\ref{phase-ja-RPA})
follows directly from the ladder structure of the RPA response functions.

From now on we adopt the $a_0=0$ gauge, in which the relation 
between $\tilde{\vec a}$ and $\vec j$ takes the form 
  \breakon
  \be
\tilde{\vec a}_{\omega,\vec k}=\frac{4\pi p i}{\omega}\,\hat z\times\vec 
j_{\omega,\vec k}
 ,\qquad
{\rm i.e.,}\qquad
\tilde{\vec a}(\vec r, t)= 4\pi p \int_{-\infty}^t \hat z\times\vec 
j(\vec r,t')dt'
 .
  \ee
  \breakoff\noindent
With this, the action $S$ finally becomes
  \be\label{phase-j*j}
S = \sum\limits_\omega \frac{2\pi p}{\omega} \left( \int 
\vec j^{\rm free}_{-\omega}(\vec r)\times\vec j_{\omega}(\vec r)\,d^2r\right)
  \ee
Note that we are working at $T=0$, and the sum over Matsubara frequencies 
should actually be interpreted as $ \int d \omega / 2 \pi $.
From the form (\ref{phase-j*j}) we proceed to evaluate $S$.

The currents $\vec j^{\rm free}_{\omega}(\vec r)$ and 
$\vec j_{\omega}(\vec r)$
are found from the diffusion and continuity equations,
  \ber
\vec j=-\hat D\nabla n
\ &,&\qquad
(\omega -\nabla\hat D\nabla) n=J_\omega(\vec r)
\ ;
\\
\vec j^{\rm free}=-\hat D^{(0)}\nabla n^{\rm free}
\ &,&\qquad
(\omega -\nabla\hat D^{(0)}\nabla) n^{\rm free}=J_\omega(\vec r)
,
\nonumber
  \eer
where $J_\omega(\vec r)=e(e^{i\omega t_1}-e^{i\omega t_2})\delta^{(2)}(\vec 
r-\vec r_0)$.
The diffusivity and resistivity tensors obey the Einstein relation
  \be
\hat D^{-1}_{\alpha\beta}=\kappa\hat\rho_{\alpha\beta}
 ,\qquad 
(\hat D^{(0)})^{-1}_{\alpha\beta}=\kappa\hat\rho^{(0)}_{\alpha\beta}
  \ee
where $\kappa$ is compressibility of free composite fermions. (Here ``free'' 
indicates
the absence of long-range interaction, whereas the short-range interaction is 
assumed to be present and to give rise to the composite Fermi-liquid physics.)

The resistivity tensors $\rho$ and $\rho^{(0)}$ are related by the composite 
fermion rule\cite{HLR}
  \be
\rho_{\alpha\beta}=\hat\rho^{(0)}_{\alpha\beta}+4\pi 
p\frac{\hbar}{e^2}\epsilon_{\alpha\beta}
  \ee
We remark that, in our notation, the diagonal tensor elements of the 
imaginary time conductivities, resistivities, and
diffusivities have a
${\rm sgn}\,\omega$ dependence on $\omega$
(see Secs. \ref{sec-model} and \ref{sec-simple} for details).  Consequently, 
we may
write $ {\hat D}(\omega) = - {\hat D}^{\rm T} (- \omega) $ and 
$n_{\omega}(\vec r)=-n_{-\omega}(\vec r)$.

Using these relations, one can simplify the expression for the action as 
follows:
  \breakon
  \ber
S &=& -\sum\limits_\omega \frac{2\pi p}{\omega}
\int d^2r 
\left(D^{(0)}_{\alpha'\alpha}\nabla_{\alpha'}n^{\rm free}_{-\omega}(\vec 
r)\right)
\epsilon_{\alpha\beta}
\left(D_{\beta\beta'}\nabla_{\beta'}n_{\omega}(\vec r)\right)
\\
&=&-\sum\limits_\omega \frac{e^2}{2h\omega}
\int d^2r 
\left(\nabla_{\alpha}n^{\rm free}_{-\omega}(\vec r)\right)
\left(\hat D^{(0)}(\hat\rho-\hat\rho^{(0)})\hat D\right)_{\alpha\beta}
\left(\nabla_{\beta}n_{\omega}(\vec r)\right)
\\
\label{S-b18}
&=&-\sum\limits_\omega \frac{e^2}{2h\omega}
\int d^2r 
\left(\nabla_{\alpha}n^{\rm free}_{-\omega}(\vec r)\right)
\left(\kappa\hat D^{(0)}-\kappa\hat D\right)_{\alpha\beta}
\left(\nabla_{\beta}n_{\omega}(\vec r)\right)
\\
\label{S-b19}
&=&-\sum\limits_\omega \frac{e^2\kappa}{2h\omega}
\int d^2r 
\left(n^{\rm free}_{-\omega}(\vec r)
\nabla\hat D\nabla n_{\omega}(\vec r)
-n_{\omega}(\vec r)
\nabla\hat D^{(0)}\nabla  n^{\rm free}_{-\omega} \right)
\\
&=&-\sum\limits_\omega \frac{e^2\kappa}{2h\omega}
\int d^2r 
\left[n^{\rm free}_{-\omega}(\vec r)
\left( -\omega +\nabla\hat D\nabla \right) n_{\omega}(\vec r)
-n_{\omega}(\vec r)
\left(-\omega + \nabla\hat D^{(0)}\nabla \right)  n^{\rm free}_{-\omega} 
\right]
\\
&=& \sum\limits_\omega \frac{e^2\kappa}{2h\omega}
\int d^2r 
\left(n^{\rm free}_{-\omega}(\vec r)J_{\omega}(\vec r)
+J_{-\omega}(\vec r)n_{\omega}(\vec r)\right)
\\
\label{phase-S-nn}
&=& \sum\limits_\omega \frac{e^2\kappa}{2h\omega}
\int d^2r 
J_{-\omega}(\vec r)
\left(n_{\omega}(\vec r)
-n^{\rm free}_{\omega}(\vec r)\right)
  \eer
  \breakoff\noindent
In the above equations, the tensors ${\hat D}$ and ${\hat D}^{(0)}$ are 
understood to
be always evaluated at frequency $\omega$, not $-\omega$. In going from 
Eq.(\ref{S-b18}) to Eq.(\ref{S-b19})
we were able to discard the boundary term because the currents normal to
the boundary are vanishing, as described below. 
The form (\ref{phase-S-nn}) will now be used for computing the action. 

The density $n_{\omega}(\vec r)$ is found by solving the diffusion equation 
in the half plane $y>0$, with the boundary condition 
$j_y=-D_{yy}\partial_y n-D_{yx}\partial_x n=0$ at $y=0$. In Fourier
components $n(x,y)=\sum_k e^{ikx}n_k(y)$ this becomes
  \ber
(\partial^2_y+q^2)n_k(y)=e(e^{i\omega t_1}-e^{i\omega t_2})\,\delta(y-y_0)
 ,
\nonumber\\
D_{yy}\partial_y n_k(y)_{y\to0}=-iD_{yx}k n_k(0)
 ,
  \eer
where $q=(k^2+\omega/D_{xx})^{1/2}$.
After solving this elementary boundary value problem we take the limit 
$y_0\to 0$ and have
  \be
n_{\omega,k}(y)=\frac{e(e^{i\omega t_1}-e^{i\omega t_2})}{D_{xx}q+i  
D_{yx}k}\,e^{-qy}
  \ee
The expression for $n^{\rm free}$ is similar, up to changing
$D_{ij}$ to $D^{(0)}_{ij}$.

By inserting $n$ and $n^{\rm free}$ thus found into Eq.(\ref{phase-S-nn})
one obtains
  \breakon
  \be\label{phase-SLH-answer}
S=\frac{e^2}{2h}\sum\limits_{\omega,k}
\,
\frac{|e^{i\omega t_1}-e^{i\omega t_2}|^2}{\omega}
\,
\left(
\frac{1}{\sigma_{xx}q+i\sigma_{yx}k}
-
\frac{1}{\sigma^{(0)}_{xx}q^{(0)}+i\sigma^{(0)}_{yx}k}
\right)
  \ee
  \breakoff\noindent
Note that this is precisely the expression for the action found in 
Ref.\cite{SLH}.
Upon evaluating the integrals over $k$ and $\omega$ it gives the result 
(\ref{Alpha_limit})
in the limit $\sigma_{xx}\to 0$ and a more general result (\ref{Alpha_limit}) 
for finite $\sigma_{xx}$.

Note that the first term in Eq.(\ref{phase-SLH-answer}), after integration over 
$k$ and $\omega$,
is a smooth function of $\sigma_{xy}$, whereas the second term gives rise to a 
cusp in the
tunneling exponent at $\sigma^{(0)}_{xy}=0$, i.e, at  $\nu=1/2$. Indeed, the 
first and the second
term of Eq.(\ref{phase-SLH-answer})
correspond to the first and the second term in Eq.(\ref{Alpha_limit}), 
respectively. This means that
the plateau in the tunneling exponent for $1/3<\nu<1/2$ arises due to the 
second term. It is
explicit in Eq.(\ref{phase-SLH-answer}) that it is the second term that accounts 
for the
free composite fermion dynamics, and so the cusp at $\nu=1/2$ should be 
understood as a signature
of the composite fermion physics. 

Let us mention, that the expression (\ref{phase-SLH-answer}) 
for the action can be rewritten as
  \be
S=\frac{e^2}{2h}
\sum\limits_\omega
\frac{\kappa}{\omega}
\Big\langle J\Bigl|
\frac{1}{\omega-\nabla\hat D\nabla}
-\frac{1}{\omega-\nabla\hat D^{(0)}\nabla}
\Bigr| J\Big\rangle
  \ee
This formula can be taken as a hint that the problem of calculating 
the semiclassical action can be significantly simplified by 
a wise choice of an effective action and of a compensating counterterm.
This is exactly what our strategy will be in the rest of the paper. 

%
%
%

\section{Effective action in $D=2$}
\label{sec-model}
\subsection{Qualitative discussion}
  Below we focus on the effect on tunneling 
arising due to relaxation of collective electrodynamical modes. 
Semiclassical theory can be used to describe it, assuming that
the times and distances controlling the tunneling rate are large.

The adequacy of the semiclassical approach can be understood as follows.
Tunneling in a strongly correlated system involves motion of a large number of 
electrons: While only one electron is actually transferred across 
the barrier, many other electrons are moving 
in a correlated fashion to accommodate 
the new electron. This collective effect becomes progressively more 
important as the bias decreases. At a small bias $V$, the single-particle 
barrier traversal time is much shorter than the relaxation 
time $\tau\sim\hbar/eV$ in the electron liquid. 
Therefore, while one electron is traversing the barrier
other electrons essentially do not move. Thus instantly 
a large electrostatic potential is formed. The jump in electrostatic 
energy by an amount much bigger than the bias $eV$  means  
that right after the one electron transfer we find the system 
in a classically forbidden state under a {\it collective} Coulomb barrier. 
In order to accomplish tunneling, the charge has to spread over 
a large area until the potential of the charge fluctuation is reduced 
below $eV$. If the conductivity is small, the spreading over a large 
distance takes a long time, and thus the action estimated as the collective 
barrier height times the relaxation time $\tau$ is much larger than $\hbar$.

This argument fully applies to a composite fermion system consisting of 
quasiparticles
interacting via Coulomb as well as Chern-Simons fields. 
The tunneling consists of an instant process of adding one electron to the 
system
and of its subsequent slow reaction. 
The second, cooperative step involving Chern-Simons and Coulomb field 
relaxation
controls the tunneling rate, while the first, single-particle step 
occurs instantly and contributes only to the 
prefactor in the tunneling current. Since for small bias the relaxation 
process
occurs on a large scale, one may describe it using the 
semiclassical approach. 
However, the fact that the tunneling particle obeys Fermi
statistics is also important, and this will be included, finally, in our 
analysis.

In what follows we treat the system motion under the collective barrier 
semiclassically as classical Coulomb and Chern-Simons electrodynamics in 
imaginary time,
find an instanton solution, and derive an expression for the tunneling rate
in terms of instanton action. For that we generalize 
to the composite fermion system the semiclassical effective 
action theory introduced elsewhere\cite{LevitovShytov}. 

\subsection{Constructing the effective action}

The effective action can be written in terms of composite fermion 
charge and current densities $n({\bf r},t)$ and
${\bf j}({\bf r},t)$, as well as the Chern-Simons gauge field $a_\mu$.
The total action is 
  \begin{equation}\label{totalS}
{\cal S}_{total}={\cal S}_{CF}+{\cal S}_{CS}+ {\cal S}_{cont}+{\cal S}_{\rm 
b.c.}
  \end{equation}
In this section we motivate, define, and discuss different parts 
of the action (\ref{totalS}) for our system. 

Below we focus on the case of diffusive CF transport taking place in
the presence of disorder. Because the electrical conductivity is local in 
this case
on scales larger than the mean free path, this problem is simpler than
that of ballistic CF dynamics. 
 
The assumption underlying our analysis is that the main contribution to the 
action
of the tunneling charge arises from large spatial and time scales, and thus
local deviation from equilibrium is small. Therefore, one can
expand the action in powers of charge and current densitites, 
$n({\bf r},t)$ and ${\bf j}({\bf r},t)$, and keep only the terms up to 
quadratic.

The contribution ${\cal S}_{CF}(n,j)$ is defined to correctly reproduce the 
equations
of motion of composite fermions decoupled from the gauge field $a_\mu$ 
but interacting via the Coulomb potential. 
[To be more precise, since composite fermions
describe interacting electrons in a magnetic field, the short-range part 
of the Coulomb interaction is included in the definition of $n$ and $j$ 
of composite fermions, so
only the residual long-range part of the Coulomb interaction enters the action 
${\cal S}_{CF}(n,j)$.] We consider ${\cal S}_{CF}(n,j)$ 
of a quadratic form constructed using CF response functions. One can see that
the requirement of matching the CF equations of motion is not entirely 
sufficient to determine
the action, e.g., because it leaves freedom of rescaling the whole 
action
or even its different parts corresponding to different normal modes of the 
problem.

The exact form of the action can be determined 
in the following way\cite{LevitovShytov}. The action used to study
the  dynamics in imaginary time is
precisely the one that appears in the quantum partition
function. The latter action expanded up to quadratic terms in
the charge and current density must yield the correct Nyquist spectrum
of equilibrium current fluctuations: 
   \begin{equation}\label{Nyquist}
\langle\!\langle {\bf g}_{\omega,q}^\alpha
{\bf g}_{-\omega,-q}^\beta \rangle\!\rangle =
\sigma_{\alpha\beta}|\omega|+
\sigma_{\alpha\alpha'}D_{\beta\beta'}q_{\alpha'}q_{\beta'} .
   \end{equation}
Here
$
{\bf g}(r)={\bf j}(r)+\widehat D\nabla n(r)
$
  is the so-called external current. In this article we are interested in
the hydrodynamical regime of small frequency $\omega$ and momentum $q$, in 
which case the
conductivity and diffusivity tensors 
$\sigma_{\alpha\beta}$ and $D_{\alpha\beta}$ satisfy the 
Einstein relation $\widehat\sigma=e^2\kappa_0 \widehat D$, where
$\kappa_0 =dn/d\mu=m^{*}/2\pi\hbar^2$ is the free CF compressibility. 
Generally, both $\widehat\sigma$
and $\widehat D$ are functions of $\omega$ and $q$.

Below we assume isotropic conductivity tensor characterized by 
$\sigma_{xx}$ and $\sigma_{xy}$. Also, to make expressions less heavy,
we often use the units $\hbar=e=1$ in intermediate steps of calculation,
and recover $\hbar$ and $e$ in final results. 

The requirement of matching equilibrium current fluctuations is
essentially equivalent to the fluctuation-dissipation theorem.
The action in imaginary time reads:
  \breakon
   \be\label{actionSn}
{\cal S}_{CF}=
{1\over 2} \sum\limits_\omega\int\!\int\! d^2\vec{r}d^2\vec{r}'\!
\left[{\bf g}_{-\omega}^{\alpha}(\vec{r})
\widehat K_{\alpha\beta}(\omega,\vec{r},\vec{r}')
{\bf g}^{\beta}_{\omega}(\vec{r}')
 +
U(\vec{r}-\vec{r}')
n_{-\omega}(\vec{r}) n_{\omega}(\vec{r}')\right]
  \ee
where $U(\vec{r}-\vec{r}')$ is the electron-electron interaction, and
the kernel $\widehat  K(\omega,\vec{r},\vec{r}')$ is  related 
to  the  current-current correlator (\ref{Nyquist}),
  \be\label{K^-1}
(K^{-1}_{\omega,q})_{\alpha\beta}
=
\langle\!\langle {\bf g}_{i\omega,q}^\alpha
{\bf g}_{-i\omega,-q}^\beta \rangle\!\rangle
=\sigma^{(0)}_{\alpha\beta}(\omega)
\omega+
\sigma^{(0)}_{\alpha\alpha'}(\omega)D^{(0)}_{\beta\beta'}(\omega)q_{\alpha'}q_{
\beta'}
  \ee
  \breakoff\noindent
   given by Eq.(\ref{Nyquist}).
Here $\widehat\sigma^{(0)}(\omega)$ and $\widehat D^{(0)}(\omega)$ are
functions of the Matsubara frequency $\omega$ obtained from
the real frequency functions by the usual analytic continuation. The 
superscript $(0)$ here
and below indicates that the response functions 
$\widehat\sigma^{(0)}$ and $\widehat D^{(0)}$ correspond to the free CF 
theory,
in the absence of coupling to the Chern-Simons field and interaction $U({\bf 
r}_1-{\bf r}_2)$.

It is appropriate to recall here the general properties of the Matsubara 
conductivity
$\sigma_{\alpha\beta} (\omega)$. By the symmetry of kinetic coefficients,
the dielectric function is an even function of the Matsubara frequency: 
$\epsilon_{\alpha\beta}(i\omega) = \epsilon_{\beta\alpha}(-i\omega)$ 
(see Ref.\cite{OnsagerPrinciple}). 
Relating it to conductivity by
$\epsilon_{\alpha\beta}(\omega) = \delta_{\alpha\beta} + 
4\pi\sigma_{\alpha\beta}(\omega)/i\omega$, 
one obtains that the longitudinal (Ohmic) conductivity
is an odd function of $\omega$, while the Hall part is an even function of 
$\omega$.
This means that the constant conductivity case actually 
corresponds to a discontinuity in $\sigma_{xx}(\omega)$ at $\omega=0$:
\be 
\sigma_{xx}(i\omega) = \sigma_{xx} \,\mathop{\rm sgn}\nolimits\omega 
\ee
whereas $\sigma_{xy}(i\omega) = \sigma_{xy}$ has no discontinuity.
The same applies to the components of the diffusivity tensor 
$D_{\alpha\beta}(\omega)$.

The coupling of composite fermion charge and current 
to the statistical gauge field  $a_{\mu}(\vec{r}, t)$ 
is described by the Chern-Simons action in a standard way \cite{HLR}:
  \be
\label{cs-1}
   {\cal S}_{CS}=i \int dt \int d^2r 
   \Big(
      na_0+\vec j\cdot\vec a
      +{1\over 4\pi p}\,  
         \varepsilon^{\mu\nu\lambda}a_\mu \partial_\nu a_\lambda
   \Big)
  \ee
Here $p$ is an even integer corresponding to the number of flux quanta
in the construction of composite fermions.

The charge and current densities entering Eqs.(\ref{actionSn})
and (\ref{cs-1}) are not independent. They may satisfy a continuity equation. 
For tunneling
problem we employ
 \be 
\label{continuity}
   \dot{n} + \nabla \vec{j} = J(\vec{r},t) \, 
 \ee 
where the source
$J(\vec{r},t)=e\delta(\vec{r}-\vec{r}_0)[\delta(t-t_1) -\delta(t-t_2)]$ 
describes adding
a composite fermion at the time $t_1$ at the point $\vec{r}_0$ and 
subsequently removing it at the
time $t_2$ at the same point. To handle this constraint, one 
has to put in the action (\ref{totalS}) the term 
  \be \label{S_continuity}
{\cal S}_{cont} = i \int \
\left(
\dot{n}(\vec{r}, t) + \nabla\vec{j}(\vec{r}, t) - J(\vec{r},t) 
\right) \, \Phi(\vec{r}, t)\ d^2r\,dt .
  \ee
with the Lagrange multiplier function $\Phi(\vec{r},t)$. 

Finally, to complete the action, one has to ensure proper boundary conditions. 
We choose the coordinates so that the 2DEG occupies the half plane $y>0$, so 
that
the half plane edge coincides with the $x$-axis. The boundary conditions at the 
edge arise
from the requirement that normal current at the edge vanishes:
  \be 
\label{boundary}
 j_{y} (x, y = 0, t) = 0
  \ee
The corresponding part of the action is constructed by
using another Lagrange multiplier:
  \be\label{S_bc}
{\cal S}_{\rm b.c.}=i\int dx \int dt\  j_y(x, y=0, t)\phi(x,t)
  \ee
Besides ensuring proper boundary conditions at $y=0$, the term (\ref{S_bc}) is 
needed to make the total action gauge invariant with respect to gauge 
transformations of the Chern-Simons field $a_\mu$. 

As remarked in Sec.~\ref{sec-phase} above, we do not 
need to include in the effective action a term expressing 
the  effect of the solenoid that appears in the
system upon the transformation of the electron
into a composite fermion.  Since we will work
in the $a_0=0$ gauge, the ``string'' phase factor
$\exp\left(i\int^{t_2}_{t_1}a_0(t')dt'\right)$
of Kim and Wen\cite{KimWen} is absent.

As a validity check of the action (\ref{totalS}) let us derive
the dynamical equations. They are obtained by taking the variation of the
action (\ref{totalS}) with respect to all variables 
excluding the Lagrange multiplier $\Phi(\vec{r}, t)$.
The resulting equations are of the standard form:
  \ber\label{EqMotion}
\widehat\rho^{(0)} \vec j    &=& \vec E_{CS} - \vec\nabla \widetilde{U} n \\
\label{EqMotion2}
{1\over 2\pi p} E^\alpha_{CS} &=& \varepsilon^{\alpha\beta}j^\beta 
                                                  \label{cs-equation}\\
\label{EqMotion3}
{1\over 2\pi p} B_{CS}   &=& n + \widetilde{J}
  \eer
where $\vec E_{CS}=\vec\nabla a_0+\dot{\vec a}$ and
$B_{CS}=\vec\nabla\times\vec a$ are Chern-Simons electric and magnetic fields. 
The effective interaction $\widetilde{U}$ is defined as
  \be\label{tildeU}
\widetilde{U}(\vec{r}-\vec{r}')=U(\vec{r}-\vec{r}')+\frac{1}{\kappa_0 
}\delta(\vec{r}-\vec{r}') ,
  \ee
where $U(\vec{r}-\vec{r}')$ is the electron-electron interaction 
and $\kappa_0=m_{*}/2\pi\hbar ^2$ is the 
compressibility of free composite fermions.
Both $\widetilde{U}$ and $\rho^{(0)}$ in Eq.(\ref{EqMotion}) in general act as 
nonlocal operators.
The boundary condition $j_y=0$, according to Eq.(\ref{cs-equation}), requires that
the tangential Chern-Simons electric field vanishes at the boundary: 
$\dot{a}_x=0$.

Also, it is straightforward to check that eliminating the
Chern-Simons field leads to Ohm's law with a corrected resistivity tensor:
\be
  \label{dynamics}
  \widehat\rho \vec j=-\vec\nabla(\widetilde{U}n) 
,\quad {\rm where}\ \
  \widehat\rho=\widehat\rho^{(0)}+{ph\over e^2}
  \left(
     \begin{array}{cc}
       0 & -1 \\
       1 & 0
     \end{array}
 \right)
\ee
is the measured resistivity tensor. 
Note that Chern-Simons interaction
changes $\rho_{xy}$, while $\rho_{xx}$ remains intact.

\subsection{The fundamental identity}
\label{sec-identity}

The nonlocal current-current term in Eq.(\ref{actionSn}) makes a calculation
for the problem in the half plane $y>0$ long and not too transparent.
To circumvent this algebraic difficulty, we derive an identity for 
the action (\ref{actionSn}) that allows us to replace it by an equivalent action 
with a local
current-current term. 

To that end, we introduce another CF action:
  \breakon
  \be\label{action-current-density}
  {\cal S}_{CF}^{\rm loc}
       ={1\over 2}\sum\limits_\omega \int d^2\vec{r}\, d^2\vec{r'}
  \left[
    {1\over \omega} 
    j^{\alpha}_{-\omega}(\vec{r}) \rho^{(0)}_{\alpha\beta}(\vec{r}, \vec{r}') 
                j^{\beta}_{\omega}(\vec{r}')
    +\widetilde U(\vec{r}-\vec{r}') n_{-\omega}(\vec{r}) n_{\omega}(\vec{r}')
  \right] ,
  \ee
  \breakoff\noindent
where $\omega$ is Matsubara frequency. 
Here $\rho^{(0)}_{\alpha\beta}(\vec{r}, \vec{r}')=
\rho^{(0)}_{\alpha\beta}\delta(\vec{r}-\vec{r}')$
is the resistivity tensor
and $\widetilde{U}$ is defined by Eq.(\ref{tildeU}).

The relation between the actions (\ref{actionSn}) and 
(\ref{action-current-density}) is provided by the following fundamental 
identity:
  \be\label{KeyIdentity}
{\cal S}_{CF}(n,j)={\cal S}_{CF}^{\rm loc}(n,j)-{\cal S}_{CF}^{\rm loc}(n_{\rm 
free},j_{\rm free}) ,
  \ee
where $n(\vec{r},t)$ and $j(\vec{r},t)$ are {\it arbitrary} functions 
satisfying the continuity equation (\ref{continuity}) and the boundary 
condition (\ref{boundary}), whereas
$n_{\rm free}(\vec{r},t)$ and $j_{\rm free}(\vec{r},t)$ correspond to the 
saddle point of the action
describing noninteracting composite fermions decoupled from 
the gauge field. Thus the functions $n_{\rm free}$ and $j_{\rm free}$
can be found by solving Eqs.~(\ref{EqMotion})--(\ref{EqMotion3}) with 
$\widetilde U(\vec{r} - \vec{r}')=\kappa_0 ^{-1}\delta(\vec{r} - \vec{r}')$ 
and no $E_{CS}$ and $B_{CS}$.
Supplemented with the continuity equation that is present in the effective 
action (\ref{totalS})
as a constraint, the equations for $n_{\rm free}$ and $j_{\rm free}$ take the 
form:
  \ber\label{EqMotion-free}
\vec j_{\rm free}(\vec{r},\omega) 
= 
- \widehat D^{(0)}\vec\nabla n_{\rm free}(\vec{r},\omega)
;
\nonumber\\
\omega n_{\rm free}(\vec{r},\omega) +\nabla j_{\rm free}(\vec{r},\omega) 
= 
J(\vec{r},\omega)
  \eer
The boundary condition for the system (\ref{EqMotion-free}) is the absence 
of normal current $j_{\rm free}$ at $y=0$.

The result (\ref{KeyIdentity}) is formulated and established below for local 
resistivity,
because in this case the proof is more straightforward. It is possible, 
however, to generalize
it to the case of nonlocal resistivity $\rho^{(0)}_{\alpha\beta}(\vec{r}, 
\vec{r}')$.
This requires more general arguments which will be discussed at the end of 
this section.

To prove the identity (\ref{KeyIdentity}), 
we write the expression (\ref{K^-1}) for the kernel $\widehat K_{\omega}^{-1}$ 
using gradients:
  \be\label{K^-1_nabla}
\left(K_{\omega}^{-1}\right)^{\alpha\beta}=\sigma^{\alpha\beta}\omega+
\left(\sigma^{\alpha\alpha'}\lvec\nabla_{\alpha'}\right)
D^{\beta\beta'}\rvec\nabla_{\beta'}
  \ee
where the operator convention is that $\rvec\nabla_{\alpha}$ acts to the 
right, whereas
$\lvec\nabla_{\alpha}$ acts to the left. 
It is useful to introduce the distinction between
$\rvec\nabla$ and $\lvec\nabla$ and to keep track of it later, so that we are 
able to invert the
kernel $\widehat K_{\omega}^{-1}$ and to evaluate the expression in the first 
term of
the action (\ref{actionSn}) before doing the integral over the half plane. 
In this way we can properly handle boundary terms. 

Inverting Eq.(\ref{K^-1_nabla}) and using the Einstein relation between
$D_{\alpha\beta}$ and $\sigma_{\alpha\beta}$ together with the relation 
between conductivity
$\sigma_{\alpha\beta}$ and resistivity $\rho_{\alpha\beta}$, one obtains
  \be
K_{\alpha\beta}=\frac{\rho_{\alpha\beta}}{\omega}-
\lvec\nabla_{\alpha}
\frac{1}{\kappa_0 \omega(\omega+\lvec\nabla\widehat D\rvec\nabla )}
\rvec\nabla_{\beta}
  \ee
Consider the first term in the action (\ref{actionSn}): 
  \breakon
  \be\label{first_term}
\vec{g}_{-\omega}^{\alpha}K_{\alpha\beta}\vec{g}_{\omega}^{\beta}
=
\frac{1}{\omega}\vec{g}_{-\omega}^{\alpha}\rho_{\alpha\beta}
\vec{g}_{\omega}^{\beta}
-\left(\nabla\cdot\vec{g}_{-\omega}\right)
\frac{1}{\kappa_0 \omega(\omega+\lvec\nabla\widehat D\rvec\nabla )}
\left(\nabla\cdot\vec{g}_{\omega}\right)
  \ee
Below we perform some manipulations with the expression (\ref{first_term}), 
refraining from integrating over $\vec{r}$ until the very end, because of the 
above-mentioned
need to be careful with gradients and boundary terms.

Now we substitute
  \be
\vec{g}_{\omega}^{\alpha}=
\vec{j}^{\alpha}+D^{\alpha\beta}(\omega)\nabla_{\beta}n
  \ee
in the first term of the right-hand side (RHS) of 
Eq.(\ref{first_term}), and find
  \be\label{exp1}
\frac{1}{\omega}\vec{g}_{-\omega}^{\alpha}\rho_{\alpha\beta}
\vec{g}_{\omega}^{\beta}
=\frac{1}{\omega}j^{\alpha}\rho_{\alpha\beta}j^{\beta}
+ \frac{1}{\kappa_0 \omega}\left(j^{\alpha}\nabla_{\alpha}n+(\nabla_{\alpha}n)j^{
\alpha}\right)
+ \frac{1}{\kappa_0 \omega}n\left(\lvec\nabla\widehat D\rvec\nabla \right)n .
  \ee
To transform the second term of the RHS of Eq.(\ref{first_term}), we substitute
  \be
\nabla\cdot\vec{g}=\nabla\cdot\vec{j}+\omega n-(\omega+\lvec\nabla\widehat 
D\rvec\nabla )n
  \ee
and obtain
  \be\label{exp2}
-\left(\nabla\cdot\vec{g}_{-\omega}\right)
\frac{1}{\kappa_0 \omega(\omega+\lvec\nabla\widehat D\rvec\nabla 
)}\left(\nabla\cdot\vec{g}_{\omega}\right)
=-J\frac{1}{\kappa_0 \omega\left(\omega+\lvec\nabla\widehat D\rvec\nabla 
\right)}J+
\frac{1}{\kappa_0 \omega}(Jn+nJ)
-
\frac{1}{\kappa_0 \omega}n\left(\omega+\lvec\nabla\widehat D\rvec\nabla 
\right)n ,
  \ee
where $J=\nabla\cdot\vec{j}+\omega n$.

Finally, we add the expressions (\ref{exp1}) and (\ref{exp2}), 
and combine the last term in Eq.(\ref{exp1}) together with
the second and third terms of Eq.(\ref{exp2}).
After doing this we find the resulting expression
  \be
\frac{1}{\omega}j^{\alpha}\rho_{\alpha\beta}j^{\beta}+\frac{1}{\kappa_0 }n^2
-J\frac{1}{\kappa_0 \omega\left(\omega+\lvec\nabla\widehat D\rvec\nabla 
\right)}J
+\frac{1}{\kappa_0 \omega}\nabla_{\alpha}\left(j^{\alpha}n+nj^{\alpha}\right)
  \ee
  \breakoff\noindent
Upon integrating this expression over $\vec{r}$ and multiplying by $1/2$, 
the first two terms give corresponding terms of the action 
(\ref{action-current-density}),
the third term gives 
${\cal S}_{CF}^{\rm loc}(n_{\rm free},j_{\rm free})$ appearing in 
(\ref{KeyIdentity}),
and the last term vanishes due to 
the boundary condition (\ref{boundary}), thus proving the identity 
(\ref{KeyIdentity})

Having given a formal proof of the identity (\ref{KeyIdentity}), let us now 
point out the
relation of Eq.(\ref{KeyIdentity}) to the structure of RPA diagrams in the 
perturbation theory
for Green's functions in the presence of disorder. To simplify the discussion, 
let us ignore
the CS gauge field, and consider the problem of electrons coupled only by 
Coulomb interaction.
In this case, the RPA self-energy $\Sigma$ can be represented graphically, as 
shown in Fig.~\ref{figRPA}.
In the $D=2$ problem the bare unscreened interaction, represented in the 
figure by a
thin broken
line, is $U(\vec{k})=2\pi e^2/\epsilon|\vec{k}|$. The diffusive polarization 
operator is
$\Pi(\vec{k},\omega)=\kappa_0  D\vec{k}^2/(\omega+D\vec{k}^2)$, and the 
diffusive vertex part
is $1/(\omega+D\vec{k}^2)$. One can verify, by performing a resummation, 
that the dynamically screened interaction, shown in Fig.~\ref{figRPA} by a 
thick black
line, can be represented as follows:
  \breakon
  \be\label{Sigma}
\frac{1}{(\omega+D\vec{k}^2)^2}\,\frac{U(\vec{k})}{1+\Pi(\vec{k},\omega)U(\vec{
k})}
=\frac{1}{\omega}\left(
\frac{1}{D\vec{k}^2+\frac{\omega}{1+\kappa_0  U(\vec{k})}} -
\frac{1}{D\vec{k}^2+\omega}\right) ,
  \ee
  \breakoff\noindent
as the difference between the propagator of an auxiliary interaction and the 
diffusive vertex part,
multiplied by $\omega^{-1}$. These two contributions are shown
in Fig.~\ref{figRPA} by the wavy red and wavy green lines, respectively.

The self-energy diagram in Fig.~\ref{figRPA} corresponds to interaction via 
a dynamically screened
Coulomb potential, i.e., to a shakeup of a dissipative plasmon. This effect 
is described by the
hydrodynamical effective action introduced above in Sec.~\ref{sec-model}, and 
so it is to be expected
that the expression in the RHS of Eq.(\ref{Sigma}) corresponds directly to the 
difference
${\cal S}^{\rm loc}-{\cal S}^{\rm loc}_{\rm free}$ in Eq.(\ref{KeyIdentity}). 

On can rewrite the formula (\ref{Sigma})
in a quite general operator form, generalizing it for any interaction 
$\widehat U$,
polarization operator $\widehat\Pi(\omega)$, and 
vertex part $\widehat V(\omega)$,
satisfying the Ward identity $\widehat\Pi(\omega)=\kappa_0  (\widehat 
1-\omega\widehat V(\omega))$.
For that, one represents the vertex part in the form $\widehat 
V(\omega)=(\widehat\lambda+\omega)^{-1}$,
and writes:
  \breakon
  \be\label{SigmaGeneral}
\widehat V(\omega)\,\left(1+\widehat U\widehat\Pi(\omega)\right)^{-1}\widehat 
U\, \widehat V(\omega)
=\frac{1}{\omega}\left(
\frac{1}{\widehat\lambda+\omega\,(\widehat 1+\kappa_0 \widehat U)^{-1}}-
\frac{1}{\widehat\lambda+\omega}
\right)
  \ee
  \breakoff\noindent 
The formula (\ref{SigmaGeneral}) is proven straightforwardly, 
by expanding the fractions in operator geometric series, and subsequent 
resummation.

\begin{figure}
\centerline{\psfig{file=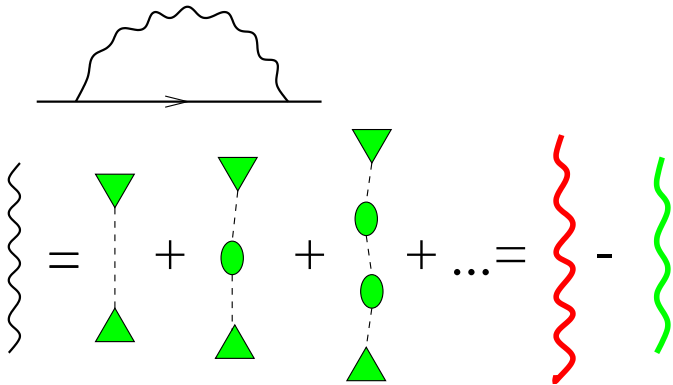,width=3.5in}}
\vspace{0.5cm}
 \caption[]{Resummation of the RPA diagram series for self-energy
   for diffusive electrons. The black wavy and broken lines represent 
dynamically
   screened interaction and bare unscreened Coulomb interaction, respectively.
   The bubbles and triangles represent diffusive polarization operators 
   and vertex parts. The red and green wavy lines
   on the RHS are defined in Eq.~\ref{Sigma}.
    }
 \label{figRPA}
\end{figure}

One can view the formulas (\ref{Sigma}) and (\ref{SigmaGeneral})
as a motivation for the identity (\ref{KeyIdentity}). More
importantly, the relation to RPA diagrams, explicit in Eq.(\ref{Sigma}) and 
(\ref{SigmaGeneral}), demonstrates the
general character of the identity (\ref{KeyIdentity}), which is not evident 
from the way
it is justified above. Comparing to Eqs.(\ref{Sigma}) and (\ref{SigmaGeneral})
makes it clear that the identity (\ref{KeyIdentity}) is robust under changes 
in the geometry of the system, alterations of the boundary conditions, 
and addition of more complicated interactions such as the CS gauge fields. 

The analog of Eq.(\ref{Sigma}) and Eq.(\ref{SigmaGeneral}),
and thus of the identity (\ref{KeyIdentity}), holds even 
for ballistic Fermi-liquid dynamics. 
In this case, according to the microscopic theory of Fermi liquid, 
$\widehat\Pi=\kappa_0  \vec{k}\vec{v}/(\vec{k}\vec{v}-\omega)$ 
and $\widehat V=1/(\omega-\vec{k}\vec{v})$, and the operators act on the 
particle-hole distributions
on the Fermi surface. For a Fermi liquid, the formula (\ref{SigmaGeneral}) 
holds with $\widehat\lambda=-\vec{k}\vec{v}$.

%
%
%
%

\section{The $D=1$ action for short-range interaction}
\label{sec-simple}
\subsection{Integrating out variables in the bulk}

In this section we consider 
the simplest model of short-range interaction,
$U(\vec{r}-\vec{r}') = U \delta (\vec{r} - \vec{r}')$, and 
diffusive CF transport described by 
$\rho_{\alpha\beta}(\vec{r},\vec{r'})=\rho_{\alpha\beta}\delta(\vec{r}-\vec{r'}
)$.

We shall start with the action ${\cal S}_{total}$ given by Eq.(\ref{totalS}) in 
the half plane
and derive an effective $D=1$ problem
by integrating out the dynamics in the bulk, and keeping only
the variables at the edge. Since the action (\ref{totalS}) is quadratic, the 
integration can easily be performed by the saddle point method. 

From now on we replace the CF action (\ref{actionSn}) by the action
(\ref{action-current-density}) with a local current-current term. The virtue of 
doing this is that
the action (\ref{action-current-density}) is much easier to handle, whereas 
the identity (\ref{KeyIdentity}) allows us to go back to the physically 
meaningful action (\ref{actionSn})
at the very end. 

First, it is convenient to integrate out the Chern-Simons gauge field 
$a_{\mu}$,
both in the bulk and at the edge. We do it by fixing the gauge $a_{0} = 0$. 
Upon integration over $a_{\mu}$ the CF resistivity tensor 
$\rho^{(0)}_{\alpha\beta}$
turns into the electron resistivity tensor  
(\ref{dynamics}): $\rho_{xy}=\rho^{(0)}_{xy}+ph/e^2$, 
$\rho_{xx}=\rho^{(0)}_{xx}$.
The action acquires the form ${\cal S}_{\rm tot}={\cal S}-{\cal S}_{\rm free}$ 
with
  \breakon
  \be 
{\cal S} = \sum\limits_{\omega}
\int{
  d^2r 
  \left(
      \frac{1}{2\omega} \, 
         j_{\alpha, -\omega} \rho_{\alpha\beta}(\omega) j_{\beta, \omega} 
    + \frac{\widetilde{U}}{2} n_{-\omega}(\vec{r}) n_{\omega}(\vec{r})    
  \right)
}
+ {\cal S}_{cont} + {\cal S}_{\rm b.c.}
  \ee
Then we integrate out $n$ and $\vec{j}$ in the bulk, keeping fixed the normal 
current $j_y$ at the edge.
The result is
  \ber 
\label{S-Phi-2D}
{\cal S} &=& \sum\limits_{\omega} 
\int{
d^2 r 
\left(
  \frac{1}{2}\,\omega\sigma_{\alpha\beta}(\omega) 
        \nabla_{\alpha} \Phi_{-\omega} (\vec{r})
        \nabla_{\beta}  \Phi_{\omega}  (\vec{r})
 +\frac{\omega^2}{2\widetilde{U}} \Phi_{-\omega} (\vec{r}) 
                              \Phi_{ \omega} (\vec{r})  
+ i\Phi(\vec{r}, t)\,  J(\vec{r}, t)\right)
}
\nonumber \\
&+& i 
\int{
   dx\, dt
      \left[
         \Phi(x, y = 0, t) - \phi(x, t)
      \right] 
      j_{y} (x, y = 0, t)  
}   
 . 
\eer
Here $\widehat{\sigma}(\omega) = \widehat{\rho}^{-1}(\omega)$ is the electron 
conductivity tensor.
The frequency dependence of $\widehat{\sigma}(\omega)$ is the same as that 
of $\widehat{\rho}(\omega)$:
$\sigma_{xx}(\omega)=\sigma_{xx}{\rm sgn}\omega$, 
$\sigma_{xy}(\omega)=\sigma_{xy}$, etc.

The next step is to integrate over $j_y(y = 0)$, which
gives $\Phi(x, y = 0, t) = \phi(x,t)$.
Hence, the action is 
  \be
\label{action-Phi}
{\cal S} = \sum\limits_{\omega} 
\int{
d^2 r 
\left(
  \frac{1}{2}\,\omega\sigma_{\alpha\beta}(\omega) 
        \nabla_{\alpha} \Phi_{-\omega} (\vec{r})
        \nabla_{\beta}  \Phi_{\omega}  (\vec{r})
 +\frac{\omega^2}{2\widetilde{U}} \Phi_{-\omega} (\vec{r}) 
                              \Phi_{ \omega} (\vec{r})  
\right)
}
+ i \int{
dx\, dt\, \Phi(x, y = 0, t) \, J(x,t)
}  . 
  \ee
  \breakoff\noindent
In handling the source term
$J$ we assume that the point $\vec{r}_0=(x_0,y_0)$ at 
which charge is injected is very close to the boundary, i.e.,
$y_0\to 0$, and thus the source in Eq.(\ref{action-Phi}) can be effectively 
placed at the
edge: $J(x, t)=e\delta(x-x_0)\left[ \delta(t-t_1)-\delta(t-t_2)\right]$.

Finally, we integrate out the bulk value of 
$\Phi(\vec{r}, t)$. From Eq.(\ref{action-Phi}) the equation for $\Phi$ at $y>0$ 
is
\be 
\label{bulk-eq}
\sigma_{xx}(\omega) \nabla^2 \Phi_{\omega}(\vec{r}) + 
\frac{\omega}{\widetilde{U}} \Phi_{\omega}(\vec{r}) = 0  . 
\ee
It is convenient to use the Fourier transform of 
$\Phi_{\omega}(\vec{r})$ with respect to variable $x$ only: 
\be 
\label{fourier}
\Phi_{\omega}(x, y) = \sum\limits_{k} \Phi_{\omega, k}(y)\, e^{ikx} .
\ee
Note that Fourier transform in $y$ is not suitable because
we are dealing with the boundary value problem in the $y>0$ domain. 

Then the solution to the equation (\ref{bulk-eq})
is straightforward: 
  \ber\label{PhiSolution}
\Phi_{\omega, k} (y) = \Phi_{\omega, k} (y = 0) e^{-q(\omega, k)\,y}
\ ; \\
q^2(\omega,k) = k^2 + \frac{|\omega|}{\widetilde{U} \sigma_{xx}}
 . \nonumber
  \eer 
After substituting Eq.(\ref{PhiSolution}) into (\ref{action-Phi}), one 
obtains a $D=1$ action: 
  \breakon
  \be 
\label{S1D}
{\cal S} = \sum\limits_{\omega, k}%
{
  \frac{1}{2}
  \left(
      \sigma_{xx} |\omega| q(\omega, k)   
    + i \sigma_{xy}\,\omega\, k         
  \right)\, 
  \phi_{-\omega, -k}\,\phi_{\omega, k} 
+ J(-\omega, -k)\,\phi_{\omega, k}   
} , 
  \ee
  \breakoff\noindent
where we put Eq.(\ref{S1D}) in the Luttinger liquid theory form in terms of the 
boundary field
$\phi(x, t) = \Phi(x, y=0, t)$ introduced above as a Lagrange 
multiplier.

This effective action represents a generalization of the
chiral Luttinger theory of edge modes to the
compressible problem with finite $\sigma_{xx}$. Because of the
relation between $q$ and $\omega$, the dissipative term in the
action (\ref{S1D}) is nonlocal in the time representation. In
the incompressible limit $\sigma_{xx}\to 0$, we recover the
standard chiral Luttinger action:
\be
  {\cal S} = {i\nu\over 4\pi}\int\,\partial_x\phi\,\partial_t\phi \,dx\,dt   
    + i\int J(x,t)\phi dx dt
\ee
In the above derivation we ignored effects of the boundary
compressibility. Taken into account, these effects lead to an
additional term of the form $\int (\partial_t\phi)^2 dx dt$
which does not affect the long-time dynamics and drops from the
final answer for the instanton action derived below.

\subsection{Instanton action}

The source term in the action (\ref{S1D}) describes coupling of the tunneling 
charge to the
field $\phi(x,t)$. Thus, the electron creation operator can be written as
$\psi^+(x,t) =\psi_{CF}^+(x,t)e^{ie\phi(x,t)}$, 
where $\psi_{CF}^+(x,t)$ is the operator of a composite fermion,
and $e$ is the electron charge. Let us point out 
the resemblance of the exponential $e^{ie\phi(x,t)}$ to the standard
one-dimensional Luttinger liquid expression.

Tunneling is related to the electron Green's function.
To find the tunneling rate, we evaluate the equal point Green's
function $G(\tau)=\langle \psi(0,t_1)\psi^{+}(0,t_2)\rangle_{\tau=t_1-t_2}$ of 
an electron.
Using the above relation of $\psi$ and $\psi_{CF}$, we write 
the electron Green's function in terms of the CF operators and then
make a {\it factorization approximation}:
  \breakon 
  \be\label{factorization}
\langle \psi(0,t_1)\psi^{+}(0,t_2)\rangle=\langle \psi_{CF}(0,t_1)\psi_{CF}^{+}
(0,t_2)\rangle
\Big\langle \exp\left(i\int J(x,t)\phi(x,t) dx dt\right)\Big\rangle  ,
  \ee
  \breakoff\noindent 
where the first and the second averages on the right-hand side 
are taken over the fermionic ground state and over fluctuations
of the electric and CS gauge fields, respectively. This approximation
holds because the dynamics of the injected quasiparticle and of the collective 
charge relaxation mode
are decoupled in space and time. The CF quasiparticles and edge magnetoplasmons 
differ both in the rate of penetration
into the 2DEG bulk and in the velocity of motion along the edge (cf. the discussion 
in Sec.~\ref{sec-phase}). 

Thus the imaginary time Green's function can be written as 
  \breakon 
  \be\label{G_Gcf}
G(\tau) = G_{CF}(\tau)
\,\Big\langle \exp\left(i\int J(x,t)\phi(x,t) dx dt\right)\Big\rangle 
= G_{CF}(\tau)\,  e^{-[S(\tau)-{\cal S}_{\rm free}(\tau)]} 
 , 
  \ee
  \breakoff\noindent 
where $G_{CF}(\tau)$ is the Green's function
of a free composite fermion injected and later removed at a point of the 
boundary.
In the last term of Eq.(\ref{G_Gcf}) we used the identity 
(\ref{KeyIdentity}) relating the average over $\phi(x,t)$ in 
$\langle \exp\left(i\int J(x,t)\phi(x,t) dx dt\right)\rangle$ to the action (\ref{S1D}).

According to the CS Fermi liquid theory, in the effective composite fermion 
mass approximation,
$G_{CF}(\tau)=1/\tau$. This essentially free fermion result holds even though 
the gauge field fluctuations
give rise to infrared-divergent logarithmic 
corrections\cite{HLR,effective-mass}
to the effective mass $m_\ast$, 
because these corrections are canceled by corrections to the residue $Z$ of 
the Green's function.

The tunneling current is obtained from $G(\tau)$ in a standard way. One has to 
continue
$G(\tau)$ from imaginary to real time, and to do the integral over time:
\be 
\label{I-V-curve}
 I(V) \sim \mathop{\rm Im}\nolimits \int_{0}^{\infty}\, 
           G(t) \, \frac{e^{ieVt}}{t} dt  .  
\ee

Now, we evaluate 
$\langle \exp\left( i\int J(x,t)\phi(x,t) dx dt\right)\rangle$ using
the {\it local} action (\ref{S1D}). By a Gaussian integration, 
the result is $e^{-S}$, where
  \be\label{Sbih}
{\cal S}={1\over 2}\sum\limits_{k,\omega} {|J(\omega)|^2\over
|\omega|(\sigma_{xx}q+i\sigma_{xy}k\ {\rm sgn}\omega) }
  \ee
The substitution $k=k_0\sinh 2x$ 
with $k_0= (|\omega|/\sigma_{xx}\widetilde U)^{1/2}$ 
simplifies integration over $k$:
  \be\label{int-S-dx}
{\cal S}={1\over 4}\sum\limits_{\omega} {|J(\omega)|^2\over
|\omega|}\,\int\frac{(e^x+1)dx}{\sigma e^x+\sigma^\ast}  ,
  \ee
where $\sigma=\sigma_{xx}+i\sigma_{xy}$. The integral (\ref{int-S-dx})
is taken in the domain $-x_{\rm max}<x<x_{\rm max}$, and 
gives an ultraviolet logarithmically divergent answer
which we cut at $k_{\rm max}=k_0 x_{\rm max}$:
  \be\label{S}
{\cal S}=\int {d\omega\over|\omega|} |J(\omega)|^2
\Big[{\rho_{xx}\over 8\pi^2} 
\ln {4k^2_{\rm max}\sigma_{xx}\widetilde{U}\over |\omega|}
+{1\over 4\pi^2}
\rho_{xy}\theta_H \Big]
  \ee
Note that this expression does not vanish even in the absence of interaction 
with
the Chern-Simons field and electron-electron interaction, when 
$p=0$ and $\widetilde U=\kappa_0 ^{-1}$. This 
indicates that part of the answer represents the contribution of
noninteracting composite fermions and must be subtracted off.
This subtraction happens automatically because of the identity 
(\ref{KeyIdentity}), which 
confirms that the correct action is indeed $S-{\cal S}_{\rm free}$. 

One can see that the counterterm ${\cal S}_{\rm free}$ is indeed related to 
the effect of free composite fermions. 
The physical origin of the ultraviolet divergence at $k_{\rm max}$ is
that for free fermions the relaxation is fast and involves large
momenta $k\sim k_F$. On the other hand, the contribution
resulting from the interaction should not diverge at large
momenta. 

To find $S-{\cal S}_{\rm free}$, we subtract from Eq.(\ref{S}) the same
expression with $p=0$ and $\widetilde{U} = \kappa_0 ^{-1}$. 
Integrating the
difference over $\omega$, we get 
$S-{\cal S}_{\rm free} = (\alpha -1)\ln t/t_0$, where
$t_0$ is a microscopic time of the order of the scattering time, and 
$\alpha$ is given by 
  \breakon 
\be
\label{Alpha}
 \alpha = 1
        + {2e^2\over\pi h}
          \Big[
              \theta_H \rho_{xy} - \theta^{(0)}_H \rho^{(0)}_{xy}
          \Big]
        + {e^2\rho_{xx}\over \pi h}
          \ln
          \Big[
             (1+\kappa_0  U) \sigma_{xx} / \sigma^{(0)}_{xx}
          \Big]
         , 
\ee
  \breakoff\noindent 
where $\theta_H=\tan^{-1}\rho_{xy}/\rho_{xx}$ is the Hall angle,
$U$ is the short-range interaction, and $\kappa_0=m_\ast/2\pi\hbar^2$ 
is the free CF compressibility.
The behavior of $\alpha$ as a function of $\rho_{xy}$ is displayed 
in Fig.~\ref{fig1}.


\begin{figure}
\centerline{\psfig{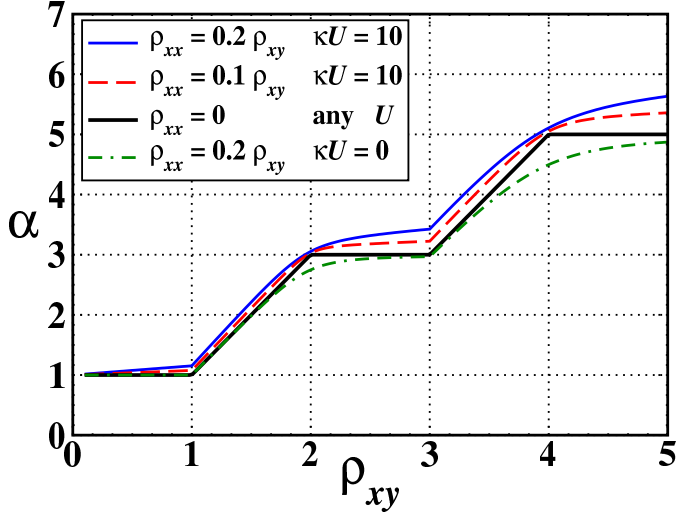}}
\vspace{0.5cm}
 \caption[]{The tunneling exponent for the model (\ref{tildeU}) 
with short-range interaction 
 $U(\vec{r}-\vec{r}') = U \delta (\vec{r} - \vec{r}')$, 
where $\kappa=\kappa_0=m_\ast/2\pi\hbar^2$.
    }
 \label{fig1}
\end{figure}

To verify that $\alpha$ is the tunneling current exponent,
we write the electron Green's function as Eq.(\ref{G_Gcf}), 
where the free composite fermion Green's function is $G_{CF}(t)\sim t^{-1}$.  
Therefore, the Green's function is $G(t) \sim t^{-\alpha}$. One can compute 
the tunneling current from Eq.(\ref{I-V-curve}), and obtain the power law
$I(V) \sim V^{\alpha}$. The expression
(\ref{Alpha}) shows that the shake-up effects suppres tunneling in a uniform 
fashion
for the filling factors $\nu$ both on and off the quantum Hall plateaus. The 
$I-V$
curve is given by a power law with the exponent depending
smoothly on the filling factor, via the resistivities
$\rho_{xx}$ and $\rho_{xy}$, and effective interaction
$\kappa_0  U$.

One can compare this result with the chiral Luttinger liquid
theories of tunneling into the edge of an incompressible
QH state. For that, one has to consider the limit of a large Hall angle:
$\theta_H=\theta^{(0)}_H=\pi/2$. In this case $\rho_{xx}\to0$ and the exponent 
(\ref{Alpha}) acquires
the form (\ref{Alpha_limit}) corresponding to a staircase with plateaus in 
the intervals
$1/3<\nu<1/2$, $1/5<\nu<1/4$, etc., interpolated by straight lines with the 
slope $2$.
At the rational filling fractions $\nu=n/(pn\pm1)$ 
we recover the results of the
Luttinger liquid theories. To see this, substitute
$\rho_{xy}=(p+1/n) h/e^2$, $\rho^{(0)}_{xy}=h/ne^2$
in the expression (\ref{Alpha_limit}), and get $I\sim
V^{1+|p+1/n|-1/|n|}$, which agrees with the universal tunneling
exponents predicted by Wen and by Kane, Fisher, and Polchinski for Jain
filling factors with positive and negative $n$.

It is interesting that the tunneling exponent (\ref{Alpha_limit})
has cusplike singularities near the compressible rational
$\nu$'s with even denominator, $\nu=1/2,\,1/4$, etc. 
The origin of this effect is a qualitative change in the 
structure of the edge modes near these filling factors. In particular, let us 
discuss the
vicinity of $\nu=1/2$, where the quantum Hall state can be
described as a Fermi liquid of composite fermions carrying $p=2$
flux quanta each, and exposed to ``residual'' magnetic field
$\delta B=(2-\nu^{-1})$. At $\nu<1/2$ the residual field
direction coincides with the total field, and all edge modes
propagate in the same direction. On the other
hand, at $\nu>1/2$, the structure of the edge is qualitatively
different, consisting
of modes going in opposite directions. This effect makes
$\nu=1/2$ a singular density from the point of view of the tunneling
exponent.

The singularities at $\nu=1/p$ are smeared in the presence of scattering by 
disorder, i.e., 
at finite $\rho_{xx}$. Interestingly, the deviation from the staircase 
described by the
expression (\ref{Alpha_limit}) due to effects of finite $\rho_{xx}$ can be 
either
positive or negative, depending on the interaction
strength $\kappa_0  U$ (see Fig.~\ref{fig1}). 
In the absence of interaction, at $U=0$, the tunneling exponent
$\alpha<\alpha(\rho_{xx}=0)$. On the other hand, at large interaction, 
$\alpha>\alpha(\rho_{xx}=0)$.

It is instructive to compare the results (\ref{Alpha}) and (\ref{Alpha_limit}) 
with the exponent $\alpha=1/\nu$ found using hydrodynamical
approaches\cite{ContiVignale,HanThouless,ZulickeMacDonald,LopezFradkin,LeeWen} 
in which the edge dynamics is modeled as a charged fluid,
without any additional inner quasiparticle degrees of freedom. 
Our expressions (\ref{Alpha}) and (\ref{Alpha_limit}) have the form of a 
difference
of two contributions, the first of which is essentially $1/\nu$ with small 
corrections due to
finite $\rho_{xx}$. The second contribution is expressed in terms of the 
response functions
of free composite fermions, and it is this term that leads to nonanalyticity 
and plateaus
in $\alpha(\nu)$. According to the identity (\ref{KeyIdentity}), these 
contributions
arise from the local action ${\cal S}_{\rm loc}$ 
and the counterterm ${\cal S}_{\rm loc}^{\rm free}$, respectively.
It is easy to see that there is a direct correspondence between our action 
${\cal S}_{\rm loc}$
and the hydrodynamical 
actions\cite{ContiVignale,HanThouless,ZulickeMacDonald,LopezFradkin,LeeWen}.
In our approach, the role of the counterterm ${\cal S}_{\rm loc}^{\rm free}$ 
is to ensure that the Green's function of free composite fermions agrees with 
Fermi statistics.
From that point of view, the plateau-like structure in $\alpha(\nu)$ is a 
manifestation
of the role of composite fermions as underlying quasiparticles of the QH 
state.

%
%
%
%

\section{Models with a long-range interaction}
\label{sec-coulomb}

\subsection{The action for the edge mode}

We assumed above that the interaction has a short range. Due to the
long-range character of the Coulomb interaction, electromagnetic modes 
in a real system are very different from those considered in 
Sec.~\ref{sec-simple}.
Hence the effect of shakeup of these modes on tunneling is also somewhat 
different.
In this section we extend the method outlined above to the problem with 
Coulomb interaction,
and consider several situations describing screening of the interaction 
in the overgrown cleaved edge system, as well as the unscreened Coulomb 
interaction\cite{fnKhvesh}.

For the long-range interaction, 
the method of deriving the effective action for the edge outlined in 
Sec.~\ref{sec-simple}
can be followed without any change up to Eq.~(\ref{S-Phi-2D}), which in this 
case takes the form
  \breakon
  \ber 
\label{S-Phi-U}
S &=& \frac{1}{2}\sum\limits_{\omega} 
\int_{y>0}
d^2 r 
\left[
  \omega\sigma_{\alpha\beta}(\omega) 
        \nabla_{\alpha} \Phi_{-\omega} (\vec{r})
        \nabla_{\beta}  \Phi_{\omega}  (\vec{r})
+ i\Phi(\vec{r}, t)\,  J(\vec{r}, t)\right]
 \\ \nonumber
&+&\frac{1}{2}\sum\limits_{\omega} 
\int\int
d^2 r d^2 r'\,\omega^2\Phi_{-\omega} (\vec{r})\widetilde{U}^{-1}(\vec{r},\vec{r
}')
                              \Phi_{ \omega} (\vec{r'})  
+ i 
\int
   dx\, dt
      \left[
         \Phi(x, y = 0, t) - \phi(x, t)
      \right] 
      \chi(x,t)
 , 
  \eer
  \breakoff\noindent 
where $\widetilde{U}^{-1}(\vec{r},\vec{r}')$ is the inverse of the interaction 
kernel,
and the notation 
  \be\label{chi-defin}
\chi(x,t)=j_{y} (x, y, t)|_{y=0}
  \ee 
is introduced. 
It will be convenient now, instead of integrating over $j_{y} (x, y = 0, t)$ 
as we did
above, to keep it as a dynamical field.

Let us note that in the interaction term in Eq.(\ref{S-Phi-U}) the integral over 
$\vec{r}$ and $\vec{r}'$ goes over the whole plane, not just over the 
half plane $y>0$
as in Sec.~\ref{sec-simple}. The reason is simple to understand by writing 
the
relation between $\Phi$ and $n$:
  \be
\omega\Phi(\vec{r})=\int_{y'>0}\widetilde{U}(\vec{r},\vec{r}')n(\vec{r}'
)d^2r' ,
  \ee
and observing that for long-range $\widetilde{U}$ the field 
$\Phi(\vec{r})\ne0$
for both $y>0$ and $y<0$. 

To proceed with deriving the effective $D=1$ action, we decompose the 
conductivity tensor
into the diagonal and off diagonal parts, 
$\sigma_{\alpha\beta}(\omega)=\sigma_{xx}{\rm sgn}\omega\,\delta_{\alpha\beta}
+\sigma_{xy}\epsilon_{\alpha\beta}$. The off diagonal conductivity 
term in Eq.(\ref{S-Phi-U}) is a full derivative, because 
 \be
\epsilon_{\alpha\beta}\nabla_{\alpha} \Phi_{-\omega} 
(\vec{r})\nabla_{\beta}\Phi_{\omega}(\vec{r})
=\nabla_{\alpha}\left[\epsilon_{\alpha\beta}
\Phi_{-\omega} (\vec{r})\nabla_{\beta}\Phi_{\omega}(\vec{r})\right] .
 \ee
As a consequence, this term is converted into the boundary term expressed in 
terms
of $\Phi_{y=0}(x,t)=\phi(x,t)$, and the total action can be written as 
  \be
{\cal S}_{\rm total}={\cal S}_{2D}+{\cal S}_{1D}
 ,
  \ee
where 
  \breakon
  \be\label{S-1D-long}
{\cal S}_{1D}=i\int \left(
  \frac{1}{2}\sigma_{xy}\partial_x\phi^{\ast}(x,t)\partial_t\phi(x,t)
+ \phi(x,t) J(x,t)
+ \left[\Phi_{y=0}(x,t)-\phi(x,t)\right]\chi(x,t)\right) dt dx
  \ee
and 
  \be\label{S-2D-short}
{\cal S}_{2D}= \frac{1}{2}\sum\limits_{\omega} 
\left(\,
\int_{y>0}
  |\omega|\sigma_{xx}
        \nabla_{\alpha} \Phi_{-\omega} (\vec{r})
        \nabla_{\alpha}  \Phi_{\omega}  (\vec{r})
d^2 r\, 
+\,\int\int
\omega^2\Phi_{-\omega} (\vec{r})\widetilde{U}^{-1}(\vec{r},\vec{r}')
                              \Phi_{ \omega} (\vec{r'})  
d^2 r d^2 r'\right)
  \ee
  \breakoff\noindent 
We included the source term $J$ in ${\cal S}_{1D}$ by placing it at the 
boundary $y=0$
and accordingly added the term $i\phi(x,t) J(x,t)$ to Eq.(\ref{S-1D-long}),
simultaneously removing the term $i\Phi_{\omega}(\vec{r}) J_{-\omega}(\vec{r})$
from Eq.(\ref{S-2D-short}).

Now, one can integrate over the field $\Phi_{\omega}(\vec{r})$. 
This amounts to taking the saddle point of 
${\cal S}_{\rm total}$, i.e., to solving the problem
  \ber\label{2D-problem}
& &-|\omega|\sigma_{xx}\nabla^2\Phi_{\omega}(\vec{r})+\omega 
n(\vec{r})=i\chi_{\omega}(x)\delta(y)
\\
& &\omega\Phi(\vec{r})=\int_{y>0}\widetilde{U}(\vec{r},\vec{r}')n(\vec{r
}')d^2r'
\nonumber
  \eer
in the domain $y>0$ with the boundary condition $\partial_y\Phi_{y=0}=0$ which 
corresponds
to the absence of  current normal to the edge. 
This problem describes the response of the charges 
in the conducting half plane 
to the external charge source $\chi_{\omega}(x)\delta(y)$. 
The solution of this problem taken at the boundary $y=0$ can be written 
as some linear operator applied to
the source $\chi_{\omega}(x)$. In terms of Fourier components one has
  \be\label{Q-def}
\Phi_{y=0}(k,\omega)=Q^{-1}(k,\omega)\frac{i\,\chi_{k,\omega}}{|\omega|\sigma_{
xx}} ,
  \ee
which defines the function $Q(k,\omega)$ playing the key role in what follows. 
Interestingly, there is no dependence 
in the problem (\ref{2D-problem}) on $\sigma_{xy}$ whatsoever, because 
the corresponding part of the action is a boundary term, and thus it belongs to
the boundary action (\ref{S-1D-long}). 

We postpone the discussion of the problem (\ref{2D-problem}) and proceed with 
deriving the
effective $D=1$ action. The integration over $\Phi_{\omega}(\vec{r})$ simply 
adds the term
$\frac{1}{2}\sum_{k,\omega} |\omega|^{-1}Q^{-1}(k,\omega)\chi_{-k,-\omega}\chi_
{k,\omega}$
to the action ${\cal S}_{1D}$ given by Eq.(\ref{S-1D-long}).

Finally, we integrate over the field $\chi(x,t)$, and obtain the total action 
in terms of the
boundary field $\phi(x,t)$:
  \breakon
  \be\label{Sloc-total}
{\cal S}=\sum\limits_{\omega,k}\,\frac{1}{2}\left[\,\sigma_{xx}\,|\omega|
\,Q(\omega,k)
+i\sigma_{xy}\,\omega\, k\, \right]\phi_{-k,-\omega}\phi_{k,\omega}+
\phi_{-k,-\omega}J_{k,\omega}
  \ee
  \breakoff\noindent 
This action, 
in which the function $Q(\omega,k)$ has to be found by solving the problem 
(\ref{2D-problem}),
represents the analog of the action (\ref{S1D}) derived in 
Sec.~\ref{sec-simple}
for short-range interaction. 

Using this action for calculating the Green's function goes 
in a complete parallel with section~\ref{sec-simple}.
The resulting Green's function is $G(\tau)=e^{-S}G^{(0)}_{\rm CF}(\tau)$,
where $G^{(0)}_{\rm CF}(\tau)=\tau^{-1}$ is the free CF Green's function. 
The saddle point action $S$, by virtue of the identity (\ref{KeyIdentity}),
can be written as ${\cal S}={\cal S}^{\rm loc}-{\cal S}^{\rm loc}_{\rm free}$,
where ${\cal S}^{\rm loc}$ and ${\cal S}^{\rm loc}_{\rm free}$ are found 
by taking an appropriate saddle point of Eq.(\ref{Sloc-total}).
The result is conveniently expressed in terms of a ``spectral weight'' ${\cal 
A}(\omega)$:
  \ber\label{formal-instanton}
G(\tau)&=&\frac{1}{\tau}\exp\left(-\int_0^\infty 
|J(\omega)|^2 {\cal A}(\omega)\,\frac{d\omega}{4\pi\hbar|\omega|}
\right)
 ,\\
J(\omega)&=&e(1-e^{-i\omega\tau}) . 
\nonumber
  \eer
Here ${\cal A}(\omega)$ is defined as
  \breakon 
  \be\label{spectral-weight}
{\cal A}(\omega)=\int_{-\infty}^\infty
\left(
\frac{1}{\sigma_{xx}Q(k,\omega)+i\sigma_{xy} k}
-\frac{1}{\sigma^{(0)}_{xx}Q^{(0)}(k,\omega)+i\sigma^{(0)}_{xy} k}
\right)\,\frac{dk}{\pi} ,
  \ee
  \breakoff\noindent 
where $Q(k,\omega)$ is defined by Eq.(\ref{Q-def}), and 
$Q^{(0)}(k,\omega)$ is determined from Eq.(\ref{2D-problem}) for 
$\widetilde{U}(\vec{r},\vec{r}')=\kappa_0 ^{-1}\delta(\vec{r}-\vec{r}')$,
which corresponds to noninteracting composite fermions. While deriving 
(\ref{spectral-weight}),
we replaced $\sigma_{xy}\omega k$ by $\sigma_{xy}|\omega|k$ in the action 
(\ref{Sloc-total}),
which does not change the integral in Eq.(\ref{spectral-weight}) because 
a sign change of $\omega$ can be accommodated by a sign change of $k$. 

The relation between the tunneling exponent $\alpha$ and the spectral weight 
${\cal A}(\omega)$
is most simple when ${\cal A}$ does not depend on $\omega$, 
as in the case of short-range interaction discussed in 
Sec.~\ref{sec-simple}. In this case,
simply $\alpha={\cal A}+1$. A frequency dependent ${\cal A}(\omega)$
can be interpreted as an energy dependent tunneling exponent 
  \begin{equation}
\alpha(\omega)={\cal A}(\omega)+1 .
  \end{equation} 
This interpretation is meaningful only if the $\omega$-dependence of ${\cal 
A}$ is sufficiently
weak. This will turn out to be precisely the case below, for the problem of 
long-range Coulomb interaction,
in which ${\cal A}(\omega)$ varies with $\omega$ not faster than 
logarithmically.

In what follows we consider the problem (\ref{2D-problem}), find 
$Q(k,\omega)$,
and evaluate the spectral weight (\ref{spectral-weight}).

\subsection{Solving for $Q(\omega,k)$}

The problem (\ref{2D-problem}) that has to be considered in order to find 
$Q(\omega,k)$
involves a long-range kernel $\widetilde{U}(\vec{r},\vec{r}')$ and, in 
general,
requires solving an integral equation. This equation is defined in the 
half plane $y>0$, and thus
cannot be treated by simple tools. 
Generally speaking, one has to treat it by the Wiener-Hopf method.

However, there are special cases corresponding to interaction 
screened by a mirror image in the region $y<0$
that can be handled by the Fourier transform. Below we consider three 
models:
  \breakon 
  \ber\label{screenedU}
{\rm model}\ V-V':&\qquad&
\widetilde{U}(\vec{r},\vec{r}')
=\frac{e^2}{\epsilon |\vec{r}-\vec{r}'|}
-\frac{e^2}{\epsilon |\vec{r}-\vec{r}''|}
+\frac{1}{\kappa_0 }\delta(\vec{r}-\vec{r}')
\ ; \\
\label{enhancedU}
{\rm model}\ V+V':&\qquad&
\widetilde{U}(\vec{r},\vec{r}')
=\frac{e^2}{\epsilon |\vec{r}-\vec{r}'|}
+\frac{e^2}{\epsilon |\vec{r}-\vec{r}''|}
+\frac{1}{\kappa_0 }\delta(\vec{r}-\vec{r}')
\ ; \\
\label{unscreenedU}
{\rm model}\ V_0:&\qquad&
\overline{U}(\vec{r}-\vec{r}')
=\frac{e^2}{\epsilon |\vec{r}-\vec{r}'|}
+\frac{1}{\kappa_0 }\delta(\vec{r}-\vec{r}')
 .
  \eer
  \breakoff\noindent 
Here the point $\vec{r}''$ is a mirror image of $\vec{r}'$ with respect to the 
edge $y=0$:
$\vec{r}'=(x',y')$, $\vec{r}''=(x',-y')$.

We start with the model $V-V'$ because it is simpler, and also because it 
directly corresponds to the
overgrown cleaved edge system where screening of the type (\ref{screenedU}) 
occurs due to the
charges induced in the doped region.
One can transform the problem (\ref{2D-problem}) in the half plane $y>0$ 
to a problem in the full plane
by extending the functions $\Phi$, $n$, and $\chi$ to the negative half plane 
$y<0$ with a sign change:
$\Phi(x,-y)=-\Phi(x,y)$, $n(x,-y)=-n(x,y)$. Similarly, the source $\chi$ in 
(\ref{2D-problem}) must be
extended so that $\chi_{\omega}(x,-y)=-\chi_{\omega}(x,y)$. 
In that, the source $\chi_{\omega}(x,y)$ is assumed to be located not right at 
the line $y=0$ but
somewhat away from it, so that the dependence of $\chi$ in 
(\ref{new-2D-problem}) below on $y$
is given by $\chi_{\omega}(x,y)=\chi_{\omega}(x)\left[\delta(y-y_0)-\delta(y+y_0)\right]$ 
with a small $y_0>0$.
The limit $y_0\to0$ will be taken at the end.

Upon extending the problem to the whole plane the interaction 
(\ref{screenedU})
has to be replaced by the unscreened interaction (\ref{unscreenedU}).
Then the problem (\ref{2D-problem}) takes the form
  \be\label{new-2D-problem}
\left(-|\omega|\sigma_{xx}\nabla^2+\omega^2\overline{U}^{-1}
\right)
\Phi_{\omega}(\vec{r})=A\delta'(y)+i\chi_{\omega}(x,y)
 ,
  \ee
where $\overline{U}^{-1}$ denotes the inverse of the operator with the kernel
(\ref{unscreenedU}). 

The term $A\delta'(y)$ is inserted because the function 
$\Phi_{\omega}(\vec{r})$, extended
from $y>0$ to $y<0$ with a sign change, must have a jump at $y=0$. The value 
of the jump
$\Phi(y=+0)-\Phi(y=-0)=-A/|\omega|\sigma_{xx}$, and thus the boundary values 
$\Phi(y=\pm0)=\mp A/2|\omega|\sigma_{xx}$. 

The formal solution of Eq.(\ref{new-2D-problem}) can be written in Fourier components:
  \be\label{Phi-Fourier}
\Phi_{\omega}(\vec{k})=
\frac{\overline{U}\left[\vec{k})(iq A+i\chi_{\omega}(k,q)\right]}{|\omega|\left[|\omega|+
\sigma_{xx}\vec{k}^2
\overline{U}(\vec{k}) \right]}
 ,
  \ee
where $\vec{k}=(k,q)$, and 
  \be\label{U-unscreened-fourier}
\overline{U}(\vec{k})=\frac{2\pi e^2}{\epsilon (q^2+k^2)^{1/2}}
+\frac{1}{\kappa_0 }
 .
  \ee
The constant $A$ is determined from the boundary condition:
  \be\label{b.c.-integral}
\partial_y\Phi_{\omega}(y\to0)=
\int iq\left(\Phi_{\omega}(\vec{k})-\frac{A}{|\omega|\sigma_{xx}iq}\right)
\frac{dq}{2\pi}
=0 ,
  \ee
where the second term in the integral is inserted to cancel the jump of $\Phi$ 
at $y=0$.

Substituting $\Phi$ from Eq.(\ref{Phi-Fourier}), 
evaluating the part of the integral (\ref{b.c.-integral}) containing 
$\chi_{\omega}(k,q)$ in the limit $y_0\to0$,
and simplifying the other part, one obtains
  \be\label{b-c-eq}
\int\frac{\chi_{\omega}(k,q)}{q}\,\frac{dq}{2\pi}
=-A\int\frac{|\omega|+\sigma_{xx}\overline{U}(\vec{k})k^2
}{
|\omega|+\sigma_{xx}\overline{U}(\vec{k})(k^2+q^2)}\,\frac{dq}{2\pi}
  \ee
Now, note that the LHS of Eq.(\ref{b-c-eq}) is equal to $i\int\chi_{\omega,k}(y) 
dy=i\chi_{\omega,k}$,
the one-dimensional source density, 
and the value of $\Phi$ at $y\to0$ 
is just given by $- A/2|\omega|\sigma_{xx}$, as discussed above. Hence, it 
follows from Eq.(\ref{b-c-eq}) that
  \be\label{Q-answer}
Q(\omega,k)=2\,\int\frac{|\omega|+\sigma_{xx}\overline{U}(\vec{k})k^2
}{
|\omega|+\sigma_{xx}\overline{U}(\vec{k})(k^2+q^2)}
\,\frac{dq}{2\pi}
  \ee
In the special case when $\overline{U}(\vec{k})$ is a constant, the result 
(\ref{Q-answer})
agrees with the expression (\ref{PhiSolution}) for $q(\omega,k)$ found in 
Sec.~\ref{sec-simple}.

The integral over $q$ in Eq.(\ref{Q-answer}) for $\overline{U}$ of the form 
(\ref{unscreenedU}), (\ref{U-unscreened-fourier})
can be evaluated exactly. We will only need the result for small $|k|\ll 
r_s^{-1}$,
where $r_s=\epsilon/2\pi\kappa_0 $ is the screening radius of the 2DEG. In 
this limit,
  \be
Q(\omega,k)=\frac{2k}{\pi}\left[\alpha\ln\left(\frac{2}{r_s|k|}\right) + 
(1-\alpha^2)F(\alpha)\right] ,
  \ee
where $\alpha=\omega\epsilon/2\pi\sigma_{xx}k$, and
  \be\label{F(alpha)}
F(\alpha)=\cases{
(1-\alpha^2)^{-1/2}\arctan\sqrt{\alpha^{-2}-1}\qquad {\rm for}\qquad \alpha<1 
\cr
(\alpha^2-1)^{-1/2}\ln\left(\alpha+\sqrt{\alpha^{2}-1}\right)\qquad {\rm 
for}\qquad \alpha>1 \cr
 }
  \ee
The expression (\ref{F(alpha)}) has no singularity 
at $\alpha=1$. The behavior of $F(\alpha)$ as a function of $\alpha$ is such 
that
$F(\alpha\ll1)=\pi/2$, $F(\alpha\gg1)=\alpha^{-1}\ln 2\alpha$, $F(1)=1$.

The next step is to substitute this expression in Eq.(\ref{formal-instanton}) to 
determine
the spectral weight ${\cal A}(\omega)$ and the instanton action. 
The resulting tunneling exponent $\alpha(\omega)={\cal A}(\omega)+1$ has a 
weak frequency dependence.
This is demonstrated on Fig.~\ref{fig-3models}, where $\alpha$ is plotted as a 
function
of frequency $\omega$ for $\nu=1/2$. 
In the two other models (\ref{enhancedU}) and (\ref{unscreenedU}), discussed 
below, the frequency dependence
of $\alpha(\omega)$ is somewhat stronger. This is quite natural because in the 
model $V-V'$
the interaction is to some extent screened by image charges, and the results 
are expected to be closer
to those for short-range interaction, where $\alpha(\omega)$ has no frequency 
dependence.
Similar difference between the effect of screened and unscreened interactions 
on tunneling is known for
the diffusive zero-bias anomaly\cite{AAL,LevitovShytov}. 


\begin{figure}
\centerline{\psfig{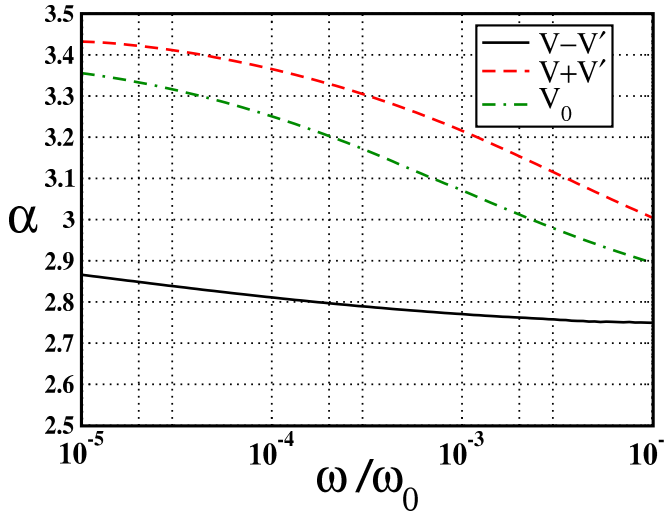}}
\vspace{0.5cm}
 \caption[]{The tunneling exponent $\alpha(\omega)={\cal A}(\omega)+1$ 
  for the models $V-V'$, $V+V'$, and $V_0$ [see Eqs.~(\ref{screenedU}), (\ref{enhancedU}), and (\ref{unscreenedU})]
  at $\nu=1/2$ as a function of frequency $\omega$. 
  The frequency is measured in units of $\omega_0=\kappa_0  e^4$. 
  For the model $V-V'$ the frequency dependence of $\alpha$
  is much weaker than for the models $V+V'$ and $V_0$. 
  Note that even in the latter two cases the frequency dependence 
  is quite weak, logarithmic at most. \\
    }
 \label{fig-3models}
\end{figure}

The model $V-V'$ is closer to the experimental situation than other models 
studied in this paper, because it treats interaction as long ranged, and 
accounts
for screening in the doped region. Thus, it is the $V-V'$ model that is 
interesting
to compare to experiment\cite{Chang2,Grayson,Chang-plateau}. 
The tunneling exponent calculated above can be plotted versus $\rho_{xy}$ (see 
Fig.~\ref{fig2}).
Experimentally, the parameter controlling occupation 
of the Landau levels is the magnetic field, and so the experimentally measured 
$\alpha$
are shown in\cite{Chang2,Grayson,Chang-plateau} as functions of $\nu^{-1}_{\rm 
bulk}=B/\Phi_0 n_{\rm 2DEG}$.
However, at large Hall angle, $\rho_{xx}\ll\rho_{xy}$, and away from 
incompressible densities,
$e^2\rho_{xy}/h$ is quite close to $\nu^{-1}$. 

Also, it would be incorrect to ignore the difference between the 
2DEG density in the bulk and near the edge, and to compare the graph in 
Fig.~\ref{fig2}
directly with the experimentally measured $\alpha$. One can argue (see 
Sec.~\ref{sec-experiment}
below) that the density near the edge exceeds $\nu_{\rm bulk}$ by $20-30\,\%$. 
Taking this into account, one has to rescale the slope of the experimentally 
observed
dependence $\alpha=1/\nu_{\rm bulk}$, and to compare the curves in 
Fig.~\ref{fig-3models}
with the dependence $\alpha=(1.2-1.3)\,\rho_{xy}e^2/h$. This agrees reasonably 
well
with the average slope of the curves in Fig.~\ref{fig-3models} in the interval
$1<\rho_{xy}<4$ studied experimentally\cite{Grayson,Chang-plateau}. 

Of course, a more important issue is whether there are plateau-like features in 
the
experimental dependence $\alpha(\nu)$. In the experiment\cite{Grayson} a 
straight line is observed,
without any sign of plateaus. More recently, however, it was found that some 
samples
show signs of a plateau near $\nu=1/3$. Upon rescaling of the filling factor 
by $1.2-1.3$,
this corresponds to $\nu_{\rm edge}$ between $1/2$ and $1/3$, which is exactly 
where the middle of the
plateau in Fig.~\ref{fig2} is located. However, the matter is clearly not yet 
resolved,
and more experimental studies would be very welcome.


\begin{figure}
\centerline{\psfig{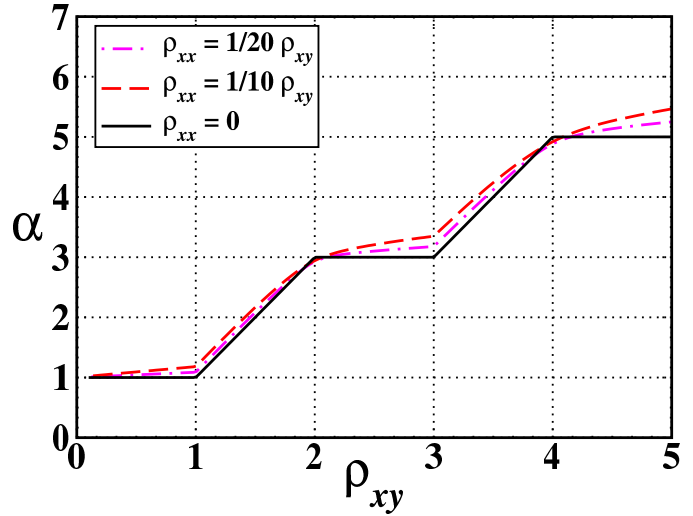}}
\vspace{0.5cm}
 \caption[]{The tunneling exponent for the model $V-V'$ 
 [Eq.(\ref{screenedU})] corresponding to Coulomb interaction 
 screened by the doped region in the overgrown edge system.\\
    }
 \label{fig2}
\end{figure}

There is one other type of interaction for which the problem 
(\ref{2D-problem}) in the half plane
is tractable by Fourier transform. It corresponds to the model $V+V'$ above, 
defined by Eq.(\ref{enhancedU}).
The interaction (\ref{enhancedU}) describes the situation when image charges 
are
of the same sign as the source charges. Despite being unphysical, this problem 
is still
worth attention, because the solution is very simple and has the behavior 
qualitatively different from
the model $V-V'$. 
Physically, this problem is similar to the one of unscreened interaction which 
we discuss
below.

Starting with the interaction (\ref{enhancedU}), one can extend the problem to 
the full
plane, now in a symmetric way: $\Phi(x,-y)=\Phi(x,y)$, etc. Upon doing this 
the
interaction (\ref{enhancedU}) has to become unscreened, 
of the form $\overline{U}(\vec{k})$ given by Eq.(\ref{unscreenedU}).
Then the solution is straightforward in Fourier components:
  \be
\Phi_{\omega}(\vec{k})=
\frac{2i\chi_{\omega,k}}{|\omega|\sigma_{xx}\vec{k}^2
+\omega^2/\overline{U}(\vec{k}) }
 .
  \ee
This form automatically satisfies the boundary condition 
$\partial_y\Phi(y=0)=0$,
because $\Phi$ is an even function of $y$. 

The function $Q(\omega,k)$ is found by evaluating $\Phi$ at the boundary $y=0$:
  \be
Q^{-1}(\omega,k)=
\frac{1}{\pi}\int\frac{dq}{q^2+k^2+|\omega|/\sigma_{xx}\overline{U}(\vec{k})}
=\frac{2}{\pi k}F(\alpha) ,
  \ee
where $F(\alpha)$ is defined by Eq.(\ref{F(alpha)}).

Again, we now substitute this expression in Eq.(\ref{formal-instanton}) to 
calculate
the instanton action. The resulting tunneling exponent $\alpha(\omega)={\cal 
A}(\omega)+1$
has a logarithmic frequency dependence, as shown in Fig.~\ref{fig-3models}. 
The  origin of this logarithmic dependence can be traced 
to the zero-bias anomaly in a diffusive conductor\cite{AAL,LevitovShytov}.
On Fig.~\ref{fig3}
we plot $\alpha$ as a function of $\rho_{xy}$ for several values of $\omega$. 
One notes
that the values $\alpha$ in Fig.~\ref{fig3} are somewhat larger than those for 
the model
$V-V'$ in Fig.~\ref{fig2}. This is due to the ``antiscreening'' in the model 
$V+V'$
which enhances the effect of the long-range part of the interaction in the 
dynamics.
Qualitatively, the behavior of $\alpha$ for the model $V+V'$ is similar to that
for the model $V_0$ discussed below.


\begin{figure}
\centerline{\psfig{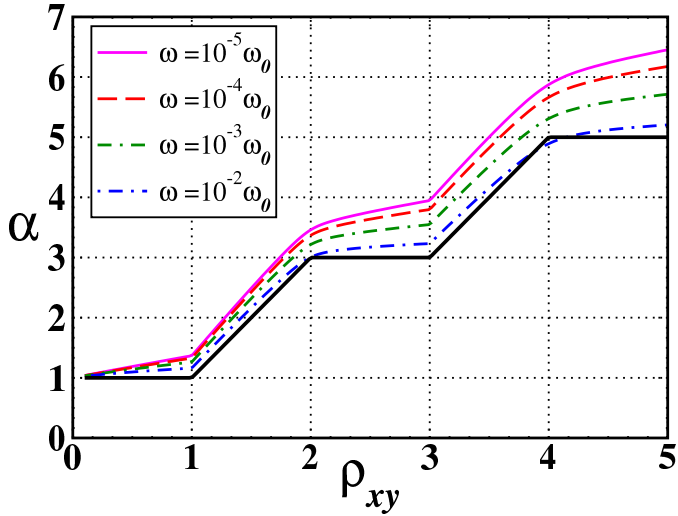}}
\vspace{0.5cm} \caption[]{The tunneling exponent for 
 the model $V+V'$ [Eq.~(\ref{enhancedU})] corresponding to
 Coulomb interaction in the presence of an ``antiscreening'' 
 due to image charges in the doped region
 of the same sign as the source charges.\\
    }
 \label{fig3}
\end{figure}

\subsection{Wiener-Hopf problem for the model $V_0$}

Here we consider the 
model $V_0$, describing unscreened Coulomb interaction 
(\ref{unscreenedU}), i.e., in the absence of 
image charges of any kind. The strategy will be to derive an integral equation 
for $\Phi_{\omega,k}(y)$ and to deal with it using the Wiener-Hopf method. 
Our approach is similar to that employed by Volkov and Mikhailov in a 
study of the edge magnetoplasmons\cite{Volkov}.

We start with the problem (\ref{2D-problem}) written in Fourier
representation with respect to $x$. Nondimensionalized, the first equation 
of (\ref{2D-problem}) reads:
 \be\label{2D-problem-nondimen}
(k^2-\partial_y^2)\Phi_{\omega,k}(y)+\mu n_{\omega,k}(y)=
i\widetilde\chi_{\omega,k}\delta(y-y_0) ,
 \ee
where $\mu=|\omega|/\sigma_{xx}$ and $\widetilde\chi=\chi/(|\omega|\sigma_{xx})
$.
As in the above discussion of the model $V-V'$,
it is convenient to place the source $\tilde\chi(y)$ at a small distance $y_0$ 
from the edge, and take the limit $y_0\to 0$ later. 

Posing the correct boundary condition for Eq.~(\ref{2D-problem-nondimen}) 
requires a discussion.
The absence of normal current at the edge means that $\partial_y\Phi(y=0)=0$.
On the other hand, by integrating Eq.(\ref{2D-problem-nondimen}) from the edge 
to the source $\widetilde\chi$, over the small interval
$0\le y\le y_0$, from current conservation one obtains 
$\partial_y\Phi(y=y_0+0)_{y_0\to0}=-i\widetilde\chi_{\omega,k}$. Therefore, in 
the limit
$y_0\to0$ the boundary condition is written as 
$\partial_y\Phi(y\to0)=-i\widetilde\chi_{\omega,k}$.
This condition defines completely the boundary value problem in the region of 
interest $y>y_0\to0$.
However, without any loss of generality, it will be convenient to assume
that near the very edge, for $0<y<y_0$, the normal derivative $\partial_y\Phi$ 
vanishes.

Now, by performing convolution of Eq.(\ref{2D-problem-nondimen}) with 
$U_k(y)=\int e^{ikx}U(x,y)dx$,
remembering that $n_k(y<0)=0$, and using the 
second equation of (\ref{2D-problem}), we transform the problem to 
  \breakon 
  \be
\label{equation-nonlocal}
\int_{y'>0}U_k(y-y')\,(k^2-\partial_{y'}^2)\Phi_{\omega,k}(y')dy'
+ \mu \Phi_{\omega,k}(y)=i\widetilde\chi_{\omega,k}\, U_k(y-y_0)
  \ee
We will be solving Eq.~(\ref{equation-nonlocal}) in the domain 
$y>0$ with $k$ and $\omega$ being parameters. Hence, for simplicity, 
below we suppress the dependence on $\omega$ and $k$ and use $\Phi(y)$, 
$U(y)$, etc.

It is convenient to integrate in Eq.(\ref{equation-nonlocal}) by parts using the 
boundary condition $\partial_y\Phi(y\to0)_{y<y_0}=0$,  which gives:
  \be
\label{equation-nonlocal-fourier}
(k^2-\partial_{y}^2)\,\int_{y'>0}U(y-y')\Phi(y')dy'
+ \mu \Phi(y)=i\widetilde\chi U(y-y_0)-\partial_{y}U(y)\Phi_{0} ,
  \ee
  \breakoff\noindent 
where $\Phi_{0}=\Phi(y=0)$. The form (\ref{equation-nonlocal-fourier})
of the problem is most suitable for applying the Wiener-Hopf method
to which we now proceed. 

The first step is to perform Fourier expansion of 
$\Phi(y)$ with respect to the $y$ coordinate: 
  \be
\label{complete-Fourier-expansion}
\Phi(y) = \sum_{q} e^{iqy}\Phi(q) 
  \ee 
Since the integral in Eq.(\ref{equation-nonlocal-fourier}) is 
taken over $y'>0$, in order to rewrite 
it in terms of $\Phi(q)$ we decompose $\Phi(y)$ as
$\Phi(y) = \Phi_>(y) + \Phi_<(y)$, 
nonzero for $y>0$ and $y<0$, respectively. One can assume that 
$\Phi_>(y)$ and $\Phi_<(y)$ decay at $y\to\pm\infty$ and verify it later,
when solution is found. In terms of $\Phi_>$ and $\Phi_<$, 
Eq.~(\ref{equation-nonlocal}) becomes
  \breakon 
  \be
\label{equation-nonlocal-Fourier}
\mu (\Phi_>(q) + \Phi_<(q)) + (k^2 + q^2) U(q) \Phi_>(q) =
iU(q)(\widetilde\chi\,e^{-iqy_0}-q\Phi_{0}) .
  \ee
  \breakoff\noindent 
Here the Fourier transformed interaction $U(q)$ is given by 
(\ref{U-unscreened-fourier}).
In what follows we set $y_0=0$.

The functions $\Phi_>(q)$ and $\Phi_<(q)$ have nice analytical properties,
namely, $\Phi_<(q)$ is an analytic function of $q$ in the upper complex 
half plane
$\mathop{{\rm Im}}\nolimits q>0$, 
and $\Phi_>(q)$ is analytic in the lower half plane $\mathop{{\rm 
Im}}\nolimits q<0$.
To make the discussion below more transparent, we denote 
$\Phi_>(q)$ by $\Phi_-(q)$, and $\Phi_<(q)$ by $\Phi_+(q)$, where $\pm$ 
indicate the
half plane of analyticity in $q$. 

Now, Eq.~(\ref{equation-nonlocal-Fourier}) can be written as
  \be 
K(q) \Phi_-(q) + \Phi_+(q) = R(q) , 
  \ee
where
  \ber\label{K-R}
K(q) &=& 1 + \frac{1}{\mu}(k^2 + q^2) U(q) \\
R(q) &=& \frac{i}{\mu}U(q)\left(\widetilde\chi-q\Phi_{0}\right)
 .
  \eer
The next step is to decompose $K(q)$ into the ratio of two
functions which are analytic in the upper and lower half planes, 
respectively: 
  \be\label{K-XX}
K(q) = \frac{X_{+}(q)}{X_-(q)}  , 
  \ee
where 
  \be
X_{\pm}(q) = 
\exp 
\left( 
       \frac{1}{2\pi i} 
        \int_{-\infty}^{\infty}
        {
             \frac{dq'}{q'-q \mp i0} \, \ln K(q')
        }
\right) .
  \ee
The asymptotic behavior of $X_\pm(q)$ at $|q|\gg2\pi\kappa_0 /\epsilon$ is 
$X_+(q)=(q+i|k|)/\lambda$, $X_-(q)=\lambda/(q-i|k|)$, where 
$\lambda=\sqrt{\mu\kappa_0 }$.

Now, Eq.~(\ref{equation-nonlocal-Fourier}) turns into
  \be
\frac{\Phi_+(q)}{X_+(q)} + \frac{\Phi_-(q)}{X_-(q)} = \Psi(q) , 
  \ee
where
  \be
\label{psi}
\Psi(q) = \frac{R(q)}{X_+(q)}
= \frac{i(\widetilde\chi-q\Phi_{0})}{q^2+k^2}\,
\left(
 \frac{1}{X_-(q)} - \frac{1}{X_+(q)} 
\right)
  \ee
Now we decompose $\Psi(q)$ into the sum of two functions with appropriate
analytical properties: 
  \breakon
  \be\label{psi-pm}
\Psi (q) = \Psi_+(q) - \Psi_-(q) 
\ ;\qquad 
\Psi_{\pm}(q) = \frac{1}{2\pi i } \int_{-\infty}^{\infty} 
\frac{dq'}{q'-q \mp i0} \, \Psi(q) .  
  \ee
The standard Wiener-Hopf reasoning\cite{Wiener-Hopf} then leads to 
  \be
\label{solution}
\Phi_+(q) = X_+(q) \Psi_+(q) \ ; \qquad
\Phi_-(q) = -  X_-(q) \Psi_-(q)  . 
  \ee
Fourier transform of Eq.(\ref{solution}) gives $\Phi(y)$ for $y<0$ and $y>0$.

It is not difficult to find $\Psi_{\pm}(q)$ explicitly. For that, one has to 
substitute Eq.(\ref{psi}) into the Cauchy integral in Eq.(\ref{psi-pm}), which gives
  \be\label{psi-integral}
\Psi_-(q)=-\frac{1}{2\pi i}\int_{-\infty}^{\infty} 
\frac{dq'}{q'-q + i0} \frac{i(\widetilde\chi-q\Phi_{0})}{q^2+k^2}\,
\left(
 \frac{1}{X_-(q)} - \frac{1}{X_+(q)} 
\right) ,
  \ee
and a similar equation for $\Psi_+(q)$. 
Now, we close the integration contour in Eq.(\ref{psi-integral}) in 
the upper or lower half plane,
depending on whether $X^{-1}_{+}$ or $X^{-1}_{-}$ is to be integrated, 
and evaluate the integral (\ref{psi-integral})
using residues. Having found $\Psi_-(q)$, and then using Eq.(\ref{solution}) 
to go back to $\Phi_-(q)$, we obtain
  \be\label{Phi-minus-final}
\Phi_-(q)=\frac{i(\widetilde\chi-q\Phi_{0})}{q^2+k^2}+
\frac{X_-(q)}{2|k|}
\left[
\frac{1}{i|k|+q}\,\frac{\widetilde\chi+i|k|\Phi_{0}}{X_-(-i|k|)}
+\frac{1}{i|k|-q}\,\frac{\widetilde\chi-i|k|\Phi_{0}}{X_+(i|k|)}
\right]
  \ee
  \breakoff\noindent 
Several remarks are in order about the result (\ref{Phi-minus-final}). 

First of all, let us verify that $\Phi_-(q)$ is analytic at 
$\mathop{{\rm Im}}\nolimits q<0$. The expression (\ref{Phi-minus-final})
has an apparent pole in the lower half plane at $q=-i|k|$. However, it is easy 
to see
from Eq.(\ref{Phi-minus-final}) that the residue 
for this pole is zero. From analyticity at $\mathop{{\rm Im}}\nolimits q<0$
it follows that $\Phi(y<0)=0$, as it should be.

Next, let us verify that the boundary value $\Phi_0$ is reproduced correctly.
For that we expand Eq.(\ref{Phi-minus-final}) in inverse powers of $q$ at 
$|q|\to\infty$:
  \be\label{Phi-series}
\Phi_-(q)=\frac{a}{iq}-\frac{b}{q^2}+...
  \ee
Since $\Phi(y<0)=0$, one simply has $\Phi(y\to+0)=a$. To evaluate $a$, only 
the first
term of Eq.(\ref{Phi-minus-final}) is important,
because $X_-(q\to\infty)=\lambda/q+O(q^{-2})$, where 
$\lambda=\sqrt{\mu\kappa_0 }$, and thus the second term
of Eq.(\ref{Phi-minus-final}) does not contribute to $a$. From the first term
one obtains $a=\Phi_0$, as expected.


\begin{figure}
\centerline{\psfig{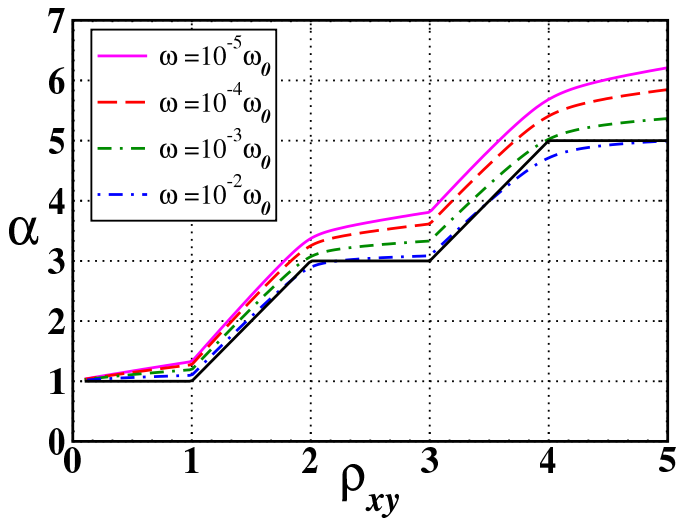}}
\vspace{0.5cm} \caption[]{The tunneling exponent for the model $V_0$
(\ref{unscreenedU}) corresponding to unscreened Coulomb interaction. 
Frequency is given in units of $\omega_0=\kappa_0  e^4$.\\
    }
 \label{fig4}
\end{figure}

After these consistency checks we can proceed with finding the relation
between $\Phi_0$ and $\widetilde\chi$. Conservation of current 
at the boundary $y=0$ for the problem (\ref{2D-problem-nondimen})
implies $\partial_y\Phi(y\to0)=-i\widetilde\chi$. On the other hand, 
$b=\partial_y\Phi(y\to0)$ in the expansion (\ref{Phi-series}).
By carrying out the expansion of the result (\ref{Phi-minus-final}) up to the 
order $q^{-2}$
to obtain $b$, and then setting up the equation $-i\widetilde\chi=b$, we 
have
  \be
-i\widetilde\chi=-i\widetilde\chi +\frac{1}{2\lambda|k|}\left[
\frac{\widetilde\chi-i|k|\Phi_0}{X_+(i|k|)}
-\frac{\widetilde\chi+i|k|\Phi_0}{X_-(-i|k|)}
\right] ,
  \ee
where $\lambda$ is the coefficient in the asymptotic expansion of 
$X_-(q\to\infty)$
defined above. This equation can be rewritten in the form
  \be\label{phi-0}
\Phi_0=\frac{i\widetilde\chi}{|k|}\,\frac{X_+(i|k|)-X_-(-i|k|)}{X_+(i|k|)+X_-(-
i|k|)}
 .
  \ee
According to Eq.(\ref{Q-def}), the relation (\ref{phi-0}) defines $Q(\omega,k)$ 
in terms of $X_+(i|k|)$ and $X_-(-i|k|)$.

The expressions for $X_{\pm}(\pm i |k|)$ can be simplified:
  \be
\label{X-pm-final}
X_{\pm} (\pm i |k|) = \exp \left[ \pm I(\omega, k)\right]  , 
  \ee
where 
  \be
\label{I-definition}
I(\omega, k)  = \frac{1}{\pi} \int_{0}^{\infty}
{ 
  \frac{d\xi}{\xi^2+1} \, 
  \ln 
  \left( 
      1 + \frac{1}{\mu} k^2(\xi^2 + 1) U(k \sqrt{\xi^2 + 1})
  \right)
}
 .
  \ee
Here $\xi = q/|k|$, $\mu=|\omega|/\sigma_{xx}$. After putting 
(\ref{X-pm-final}) into
(\ref{phi-0}), one finally arrives at
  \be
\label{Q(w,k)-final}
Q(\omega,k)= |k|\coth I(\omega, k)
  \ee
With this expression for $Q(\omega,k)$ one can go back to the effective 
action (\ref{Sloc-total}), and find the Green's function 
(\ref{formal-instanton})
in terms of the spectral weight ${\cal A}(\omega)$ given by 
(\ref{spectral-weight}).

The integral entering Eq.(\ref{I-definition}) can easily be tabulated 
numerically.
The spectral weight ${\cal A}(\omega)$ has a logarithmic frequency dependence, 
as shown in Fig.~\ref{fig-3models}, similar to that of the model $V+V'$. 
The behavior of the tunneling exponent $\alpha$ as a function of $\rho_{xy}$, 
shown in Fig.~\ref{fig4}, is also close to that for the model $V+V'$. 
One notes that the values of $\alpha$ are somewhat less than those for the 
model $V+V'$
with similar parameters. 
This is due to a relatively weaker effect of the long-range part of the 
interaction
in the model $V_0$.

%
%
%
%
%
%

\section{Comparison to the experiment}
\label{sec-experiment}

In this section we discuss some aspects of the overgrown 
cleaved edge system\cite{Chang2,Grayson,Chang-plateau}.
In our view, the most relevant issue concerns the 2DEG density distribution 
near the edge.
One of the key features of cleaved edge systems is that
they produce structures with supposedly an atomically sharp confining 
potential,
and thus the 2DEG density profile near the edge is expected to be reasonably 
smooth.
This is important in edge tunneling experiment, because 
the system must have a well defined filling factor even very close to the 
edge.

\subsection{Thomas-Fermi model}

To estimate the importance of various factors controlling the density near the 
edge,
below we consider a simplified electrostatic Thomas-Fermi model, in which the 
2DEG is modeled
as an ideal charge fluid, and all effects of electron-electron correlation 
and finite density of states are 
ignored, 
except very close to the edge. 
In principle, this approximation is quite reliable at distances larger 
than the screening length $r_s=\epsilon/2\pi\kappa_0 $, and so the results 
will be meaningful at distances more than $r_s$ from the edge. 

The electrostatic problem we consider involves the 2DEG density $n(x,y)$ in 
the half plane $y>0$,
top surface charge states that are at a distance $w=600\,{\rm nm}$ above the 
2DEG,
a layer of charged donors parallel to the 2DEG at a distance $w_+=60\,{\rm 
nm}$
above the 2DEG plane, and also charges in the three-dimensional doped region, 
which in our model occupies the halfspace $y<-w_b$, where $w_b=9\,{\rm nm}$ 
is the width of the barrier together with the buffer region. 
The top surface, the 2DEG, and the doped region are assumed to be 
equipotentials in the problem. 
For simplicity, we assume that the 2DEG is grounded, and the bias voltage on 
the 3D doped region is very small, so that
the electrochemical potentials of the two regions are essentially equal.
Relative to the 2DEG, the electrostatic potential at 
the top surface is $V_s=-800\,{\rm mV}$, and the 
electrostatic potential at the boundary of the 3D doped region is 
$V_d\approx 20\,{\rm mV}$. 
(The value of $V_d$ reflects the chemical potential difference 
before the charge redistributes itself. It is given by the
difference of Fermi energies in the doped region and in the 2DEG
plus the confinement energy of the 2DEG.)  
The charge density of donors $\sigma_+$ is taken to be constant everywhere 
at $y>0$ up to the edge $y=0$. The potential $V_d$ is much smaller than the 
barrier height, which is estimated as $\simeq 120\,{\rm meV}$.

One can write down a simple analytic formula for the 2DEG density, using the 
electrostatic
superposition principle, according to which the effects on the 2DEG due to the 
donors, the top surface charge, and the doped region, can be treated 
separately and then added.

First, let us consider the charge induced by donors, when the top surface 
and the doped region are at the same electrostatic potential as the 2DEG. We 
make
an approximation $w\gg w_+$, which allows us to move the top surface 
to infinity, and thus to ignore it. Also, we assume that the distance to the 
doped region $w_b\ll w_+$, the separation of the donors from the 2DEG. 
With the values for $w$, $w_+$, and $w_b$
quoted above, both approximations are reasonable. The resulting 
contribution to the 2DEG charge density is
  \be
\sigma^{(1)}_{\rm 2DEG}(y)=\frac{2\sigma_+}{\pi}\,\arctan\,\frac{y}{w_+} .
  \ee
It describes the 2DEG density, constant and equal to $\sigma_+$ at $y\gg w_+$, 
and decreasing to $0$ near the edge. 

The effect of the top surface potential $V_s$, in the absence of donors, and 
with
the 2DEG and the doped region at zero electrostatic potential, can be 
evaluated as
follows. In the approximation
$w_b\ll w$, the problem is equivalent to the standard electrostatic problem of 
a half-open slit,
with one side of the slit being at the potential $V_s$ with respect to the 
other side
and the end. The induced charge density in this problem is
  \be
\sigma^{(2)}_{\rm 2DEG}(y)=\frac{\epsilon V_s}{4\pi w}\,\tanh\,\frac{\pi 
y}{2w} .
  \ee
This contribution is constant and equal to $\epsilon V_s/4\pi w$ in the bulk, 
at $y\gg w/\pi$,
and decreases to zero near the edge. 

Finally, the effect of potential difference between the 2DEG and the doped 
region
can be considered ignoring the top surface and the donors. The relevant 
spatial scale
in this case is $w_b\ll w_+,w$, and so the problem is reduced to that of 
a ground half plane (representing the 2DEG), and a conducting plane 
perpendicular to it, at a relative potential $V_d$, located a distance $w_b$ 
away from the
ground half plane. The charge density induced in the 2DEG is
  \be
\sigma^{(3)}_{\rm 2DEG}(y)=\frac{\epsilon V_d}{2\pi^2}\,\frac{1}{\sqrt{(y+w_b)^
2-w_b^2}} .
  \ee
It behaves as $1/y$ away from the edge, and as $1/\sqrt{y}$ near the edge. 
The square root divergence near the edge is an artifact of the simplified model
ignoring finite density of states of the 2DEG. In a Thomas-Fermi model, 
the divergence would 
be cut at a distance $\sim r_s$ from the edge. 


\begin{figure}
\centerline{\psfig{file=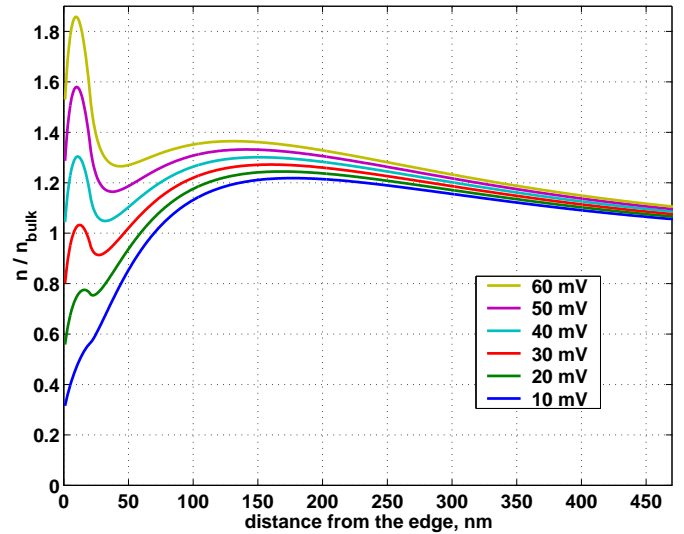,width=3.5in}}
\vspace{0.5cm} \caption[]{Density distribution in the 2DEG near the 
edge plotted for six values of the potential $V_d$ of the doped region 
(listed from top to bottom). 
The top surface potential $V_s=-800\,{\rm mV}$ and the donor density
$\sigma_+=1.94\cdot 10^{11}\,{\rm cm}^{-2}$ 
correspond to the 2DEG bulk density $n_{\rm bulk}=10^{11}\,{\rm cm}^{-2}$. 
The geometrical parameters used are defined in the text.
    }
 \label{fig-2deg}
\end{figure}

The resulting 2DEG charge density is a sum of three terms, 
$\sigma_{\rm total}=\sigma^{(1)}_{\rm 2DEG}+\sigma^{(2)}_{\rm 
2DEG}+\sigma^{(3)}_{\rm 2DEG}$.
To eliminate the unphysical singularity near the edge due to $\sigma^{(3)}$,
we average the density $\sigma_{\rm total}$ over intervals of length $2r_s$, 
and consider
  \be
\sigma_{\rm total}^{\rm av}(y)=(2r_s)^{-1} \int^{y+r_s}_{y-r_s}\sigma_{\rm 
total}(y')dy' .
  \ee 
The averaged density $\sigma_{\rm total}^{\rm ave}$ is plotted in 
Fig.~\ref{fig-2deg}
for several values of the doped region potential $V_d$. The screening radius 
used in the
averaging is taken to be $r_s=20\,{\rm nm}$. 

One can see from Fig.~\ref{fig-2deg} that the density within $\simeq 200\,{\rm 
nm}$
near the edge is quite sensitive to the potential $V_d$. Another feature 
evident in
Fig.~\ref{fig-2deg} is that the density close to the edge exceeds that in the 
bulk by $20-30\,\%$.
The 2DEG density approaches the bulk value at distances $\ge 400\,{\rm nm}$ 
from the edge.
Also, there is a peak in the density profile near the very edge, resulting 
from the $\sigma^{(3)}$
contribution averaged over the length $\simeq r_s$. This peak makes the 
density profile nonmonotonic, with a minimum at $\simeq 30-40\,{\rm nm}$ from 
the edge.
Altogether, the 2DEG density near the edge is smooth but not perfectly 
uniform.

It should be remarked that our simplified electrostatic model is perhaps 
insufficient
at distances smaller than or of the order of $r_s\simeq 20\,{\rm nm}$. Thus the
smallest scale features in Fig.~\ref{fig-2deg}, such as the density peak near 
the edge,
should be taken with caution. Moreover, we used the Thomas-Fermi model, the 
screening radius $r_s$, etc.,
in the absence of magnetic field. It remains to be seen whether the 
results
are preserved in a more accurate treatment accounting
for Landau levels, finite 2DEG compressibility, and exchange effects. 
On the other hand, on spatial scales larger than $r_s$, 
the results obtained from a purely electrostatic model should be reliable. 

One issue that can be addressed using the electrostatic model is the 
calibration of density
in the experiment \cite{Chang2,Grayson}. The tunneling exponent $\alpha$ is 
presented there
as a function of magnetic field, which is calibrated in terms of the bulk 
filling factor
using magnetotransport data. However, the filling factor relevant for 
tunneling is that
near the edge. According to the above, 
in the region 100 $\sim$ 300 nm from the edge, the density
is at least $20-30\%$ higher than in the bulk. 
If one assumes that this is the relevant distance scale for charge relaxation 
at
the temperatures and voltages employed in the experiments, then 
the dependence $\alpha=1/\nu_{\rm bulk}$ observed in 
\cite{Grayson,Chang-plateau}
translates into
$\alpha\approx (1.2-1.3)\,\nu^{-1}_{\rm edge}$. In actuality, 
the relevant distance scale
will depend on the filling factor and the cleanliness of the edge, as well as 
the
energy of the tunneling electron.

One notes that after accounting for the 
difference between  $\nu_{\rm edge}$ and 
$\nu_{\rm bulk}$ the dependence $\alpha(\nu)$ shifts 
closer to the theoretical curves (see Fig.~\ref{fig-comp-edge}).

\subsection{Two-mode model}

Because the 2DEG density profile discussed above is 
significantly nonmonotonic near the edge, 
it is possible that this may change the structure of the edge modes. 
More precisely, suppose that the
peak density near the edge is so high that the filling factor reaches $\nu=1$ 
within the
region $\simeq 30\,{\rm nm}$ corresponding to the peak displayed in 
Fig.~\ref{fig-2deg}.
Then the edge mode on the periphery will correspond to $\nu=1$ even when 
$\nu<1$ away from the edge.
In this case, in addition, there will also be counterpropagating modes 
positioned on the inner side
of the incompressible $\nu=1$ region. The number of these modes and their 
Hamiltonian will depend
on $\nu$ somewhat away from the edge. This type of acomposite structure of 
the edge was first
proposed by MacDonald for the $\nu=2/3$ system, based on a Hartree-Fock 
analysis\cite{composite-edge}.

In this model, the tunneling electron is injected into the outer $\nu=1$ mode, 
because of higher overlap of the tunneling state with the mode closest to the 
edge.
We assume that the edge is so clean that we can neglect scattering between 
different
edge modes. Then, 
the inner modes will be important only to the extent
that tunneling charge couples with them by Coulomb interaction, and shakes 
them up.
In this scenario, after tunneling there is no statistics change of the 
injected particle,
since it remains in the fermionic $\nu=1$ edge state. Therefore, one expects a 
smooth
dependence of the tunneling exponent on $\nu$, without any cusps or plateaus. 

To estimate the shakeup effect due to Coulomb coupling to the inner modes, 
let us represent them by a single charged mode.  
Thus the system can be described by two counterpropagating chiral modes:
  \breakon
  \be\label{S-2-modes}
{\cal S}=\frac{1}{2}\sum\limits_{\omega,k}\left(
i\omega k\,\left( \phi^{(1)}\phi^{(1)}
-g\phi^{(2)}\phi^{(2)}\right)
+\omega^2\sum\limits_{i,j=1,2}V_{ij}\phi^{(i)}\phi^{(j)}
\right)+\phi^{(1)}J ,
  \ee
  \breakoff\noindent 
where $g=1-\nu$ and $V_{ij}$ is the coupling matrix, expressed in terms of the 
interactions $V_{ij}^{0}$
as follows: $V_{11}=V_{11}^{0}$, $V_{12}=g V_{12}^{0}$, 
$V_{21}=g V_{21}^{0}$, $V_{22}=g^2 V_{22}^{0}$. The form of the action 
(\ref{S-2-modes})
can easily be justified in the same way as in Sec.~\ref{sec-simple}. 
In this case there is no issue of charge injection in the inner mode, and so 
there are no complications related with counterterms, as in 
(\ref{KeyIdentity}).

It is straightforward to write down the Green's function by evaluating the 
saddle point
of the quadratic action (\ref{S-2-modes}). The result reads
  \breakon
  \be
G(\tau)=\exp\left(-\frac{1}{8\pi^2}\int
\frac{(\omega^2 V_{22}-ig \omega k)|J_{\omega}|^2\, d\omega\,dk}
{(\omega^2 V_{11}+i\omega k)(\omega^2 V_{22}-ig \omega k)-\omega^4 
V_{12}V_{21} }
\right)
  \ee
  \breakoff\noindent 
To evaluate the Green's function, we assume that the coupling matrix $V_{ij}$ 
has no
$k$ dependence. This is true for the screened Coulomb interaction $2\pi 
e^2(1-e^{-2a|k|})/\epsilon |k|$
at $a|k|\le 1$, where $a$ is the distance from the edge mode location to the 
doped region.
Hence the length $a$ is somewhat larger than the barrier width $w_b$. 

In this case the integral over $k$ can be done by residues, and the result is 
$G(\tau)=\tau^{-\alpha}$,
where 
  \be\label{comp-edge-alpha}
\alpha=\frac{V_{11}^{0}+g V_{22}^{0}}{\sqrt{(V_{11}^{0}+g 
V_{22}^{0})^2-4gV_{12}^{0}V_{21}^{0}}}
  \ee
The dependence $\alpha(\nu)$ in the interval $0<\nu<1$ is smooth, 
without singularities, as it should be in the case when the effect of the 
fractional edge
is purely a shakeup, not accompanied by injection of charge.

To estimate numerical values of $\alpha$, we consider 
a model in which the interactions $V_{ij}^{(0)}$ are given by the Coulomb 
potential
screened by the doped region. We assume that the outer edge is separated from 
the doped region
by a barrier of thickness $w$, and the inner and outer edge states are a 
distance $a$ apart.
Then 
  \breakon 
  \be
V_{11}^{(0)}=\frac{2\pi e^2}{\epsilon |k|}(1-e^{-2w|k|}),\qquad
V_{22}^{(0)}=\frac{2\pi e^2}{\epsilon |k|}(1-e^{-2(w+a)|k|}),\qquad 
V_{12}^{(0)}=V_{21}^{(0)}=\frac{2\pi e^2}{\epsilon |k|}e^{-a|k|}(1-e^{-2w|k|})
\nonumber
  \ee
  \breakoff\noindent 
We consider the limit of small $k$, where the interactions $V_{ij}^{(0)}$
do not depend on $k$: $V_{11}^{(0)}=2w(2\pi e^2/\epsilon)$,
$V_{22}^{(0)}=2(w+a)(2\pi e^2/\epsilon)$, 
$V_{12}^{(0)}=V_{21}^{(0)}=(2w-a)(2\pi e^2/\epsilon)$. 

\begin{figure}
\centerline{\psfig{file=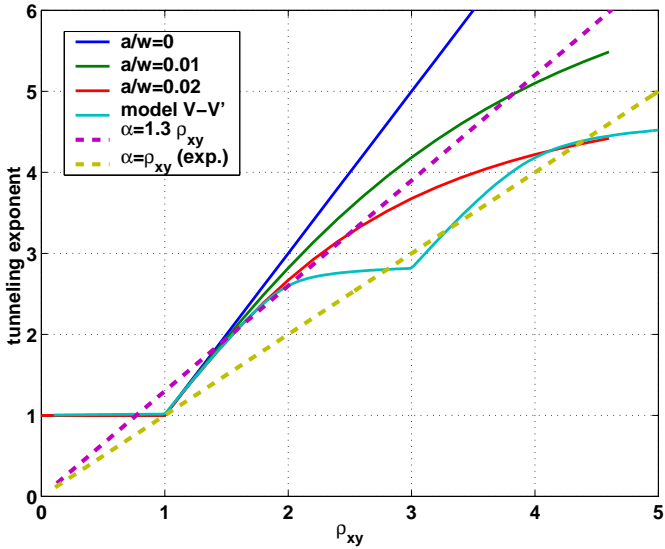,width=3.5in}}
\vspace{0.5cm} \caption[]{
{\it Solid curves:\ }
The tunneling exponent in the composite edge model (\ref{S-2-modes}) 
is shown for 
three values of the
ratio $a/w$ of the distance between the outer and inner edges and the 
tunneling barrier width.
For comparison, a theoretical curve for the $V-V'$ model is shown, for 
$\rho_{xx}=\rho_{xy}/10$ and $\omega=10^{-5}\kappa_0  e^4$.
{\it Dashed curves:\ }
The straight line $\alpha=\rho_{xy}$ corresponds to  
experiment (Refs.~\cite{Chang2,Grayson,Chang-plateau}), and the
line $\alpha=1.3 \rho_{xy}$ is obtained by correcting the filling factor by 
the ratio of the densities
near the edge and in the bulk.

    }
 \label{fig-comp-edge}
\end{figure}

In this model, the only parameter is the ratio $a/w$. The tunneling exponent 
is plotted
in Fig.~\ref{fig-comp-edge} as a function of $\rho_{xy}=\nu^{-1}$ for several 
values
of $a/w$. On the same figure, we show the experimental dependence of $\alpha$ 
versus
$\rho_{xy}$ rescaled by a factor $1.3$ as discussed above. 

The distinct feature of the composite edge model
is the absence of plateaus in the tunneling exponent $\alpha(\nu)$. However,
note that in order for the tunneling exponent $\alpha$ 
to fall in the right range, one has to assume unphysically small values of the 
ratio $a/w$.
Also, the theoretical curves for nonzero $a/w$ have  curvature 
which is absent in the experimental curve. This curvature is even more 
significant at higher values of the parameter $a/w$ and is unlikely to 
disappear if
one takes into account possible dependence of $a/w$ on $\nu$.
It is apparent that this simplified two-mode model
does not agree 
with the experimental results on tunneling. Nevertheless, it illustrates the 
point that, if scattering between edge modes is sufficiently small, 
a complicated edge
structure can lead to large changes in the observed tunneling exponent, which 
will not
be closely related to the bulk filling factor.

\section{Summary}

The problem of tunneling into the edge of a composite fermion QH system is 
treated
for long-range Coulomb interaction between electrons,
as well as for a short-range interaction model.
It is shown that in the case of diffusive CF dynamics described by finite 
$\rho_{xx}$,
the tunneling exponent is controlled by 
the coupling of tunneling electron to the charged edge mode. The effective 
action for this mode
is a generalized chiral Luttinger action with a nonlocal dissipative term. 

The tunneling exponent is found to be a continuous 
and monotonic function of $\rho_{xy}$, given, in the limit
$\rho_{xx}\to 0$,  by $\alpha=1+\frac{e^2}{h}(\rho_{xy}-|\rho^{(0)}_{xy}|)$, 
where $\rho^{(0)}_{xy}$ is the CF Hall resistivity due to motion in the 
residual magnetic field.
In order to verify the robustness of the results we consider several models for
the electron-electron interaction: the short-range and Coulomb interaction, 
and,
in the latter case, with and without electrostatic screening due to image 
charges in the doped region.

The dependence of $\alpha$ on $\rho_{xy}$ is characterized by plateaulike 
features,
not observed in the experiments on cleaved edge systems. We discuss the 2DEG 
density
profile near the cleaved edge, and propose that the discrepancy between  
theory and
experiment is possibly due to spatial variation of the density near the edge and, 
in particular,
to a nonmonotonic density profile, giving rise to 
a composite structure of edge modes.

\vskip5mm
\noindent
{\bf Acknowledgments:} \\
We are grateful to Albert Chang and Matt Grayson for many useful discussions 
and for
sharing their data prior to publication. We thank Bell Labs for support
and hospitality at the initial stage of this project.
This work was supported in part by the MRSEC Program 
of the NSF
under Grant No. DMR 98-08941 and by NSF Grant No. DMR 99-81283.

\vskip5mm

\appendix

\section{Estimate for the ballistic regime}  
\label{ballistic}

In the ballistic regime, for length scales smaller than the composite
fermion mean free path $l$, the conductivity tensor is nonlocal in space.
Close to the edge, the CF conductivity $\sigma^{(0)}_{\alpha \beta}
(\bf{r}, \bf{r^\prime}, \omega)$ depends on the distance from the edge, as
well as on the separation $\bf{r} - \bf{r^\prime}$.  As a crude
approximation, however, in order to estimate the contribution of the
short-distance response to the tunneling exponent, we shall ignore the
dependence on the distance from the edge and use, instead, the bulk CF
conductivity, which depends only on $\bf{r} - \bf{r^\prime}$~\cite{fnKhvesh}.

As discussed in Secs.~\ref{sec-model} and~\ref{sec-simple}, 
for a nonlocal conductivity we may still
approximate the Green's function $G (\tau)$ using the factorization
(\ref{G_Gcf}), but the actions $S(\tau)$ and $S_{\rm free}(\tau)$ should be
evaluated using the correct nonlocal conductivity.  Instead of this, in 
our approximation, we use the form (\ref{Sbih}) for ${\cal S}$, with the
change that we replace the macroscopic conductivity $\sigma_{xx}$ by the 
quantity $\sigma_l (|k|)$, which is the wave-vector dependent
longitudinal conductivity for the bulk compressible Hall state.
Specifically, at $\nu = 1/2$,  according to Ref.\cite{HLR}, we have   
  \be\label{xb1}
\sigma_l (k) \sim (e^2 / 8 \pi \hbar) (k / k_F) ,
  \ee
for $l^{-1} < k < k_F$, while $ \sigma_l (k)$ reduces to the 
macroscopic conductivity $\sigma_{xx}$ for $k < l^{-1}$.
We continue to approximate the Hall conductivity in Eq.(\ref{Sbih}) by its
macroscopic value $\sigma_{xy}$.

In evaluating Eq.(\ref{Sbih}), it is convenient to combine contributions from
wave vectors $k$ and $-k$, and replace the sum over $k$ by an integral over
positive values of $k$.  If the frequency $\omega$ is sufficiently small,
there will be two distinct regions which can contribute significantly to
the integral.  The first, for $k<l^{-1} $, gives the same 
contribution as was
found in Sec. \ref{sec-simple} above, since  $\sigma_l (k) = \sigma_{xx}$ in this
region.  In the region $k> l^{-1} $, we may set $q=k$, when $\omega$ is
sufficiently small, so that the integrand takes the form 
\be\label{xb2}
 I_{\rm large k} = {|J(\omega)|^2\over
|\omega| |k| \sigma_{xy}} {\rm Re} 
{\sigma_{xy} \over (\sigma_l(k)+i\sigma_{xy}) } .
  \ee.

The last factor in Eq.(\ref{xb2}) is small for $k \ll k_F$ 
but becomes of order unity for 
$k \approx k_F$, where $\sigma_l (k) \approx \sigma_{xy}$. The contribution
from this region to the integral could therefore make a contribution of
order unity to the
tunneling exponent $\alpha$.  However, this contribution may be largely or
completely canceled by the corresponding contribution to $S_{\rm free}$.

If we neglect the difference between the longitudinal and transverse
conductivities at the finite wave vector $k$, then 
  \be\label{xb3}
{\rm Re} {1 \over (\sigma_l(k)+i\sigma_{xy}) } = \rho_l (k)
  \ee
which is the longitudinal resistivity at wave vector $k$.  In calculating
$S_{\rm free}$, using the same assumptions, we obtain the identical expression,
because the longitudinal conductivity of the composite fermions is the same
as that of the electrons.  Thus the contribution to the tunneling exponent
from short wavelengths is canceled, in this approximation.  We therefore
wind up with the same value for $\alpha$ as was obtained in 
Sec.~\ref{sec-simple},
namely $\alpha \sim 3$ at $\nu = 1/2$, for a system where $\sigma_{xx} \ll
\sigma_{xy}$.

It is not possible to say whether a similar cancelation would occur in a
proper analysis incorporating the nonlocal conductivity. If the cancellation
does not occur, then the surviving contribution from short wavelengths
could give a contribution of order unity to the tunneling exponent, 
which would be independent of the mean-free-path 
in the limit $l \ll k_F^{-1}$.

\end{multicols}

\begin{thebibliography}{99}

\bibitem{Wen} 
  X.-G.~Wen, {\sl Int. J. of Mod. Phys.} {\bf B6}, 1711 (1992);
  {\sl Phys. Rev. Lett.} {\bf 64}, 2206 (1990);
  {\sl Phys. Rev.} {\bf B43}, 11025 (1991);
  {\sl Phys. Rev.} {\bf B44}, 5708 (1991);

\bibitem{Fisher}
  C.L.~Kane, M.P.A.~Fisher and J.~Polchinski,
  {\sl Phys. Rev. Lett.} {\bf 72}, 4129 (1994);
  C.L.~Kane and M.P.A.~Fisher, {\sl Phys. Rev.} {\bf B51}, 13449 (1995)

\bibitem{Webb}
  F.P.~Milliken, C.P.~Umbach, and R.A.~Webb, 
       {\sl Solid State Comm.} {\bf 97}, 309 (1996);

\bibitem{Pfeiffer'90} 
  L.N.~Pfeiffer, K. W. West, H. L. Stormer, J. P. Eisenstein, K. W. Baldwin, D. Gershoni, 
 and J. Spector,
{\sl Appl. Phys. Lett.} {\bf 56}, 1697 (1990)

\bibitem{Chang2}
  A.M.~Chang, L.N.~Pfeiffer, and K.W.~West,
       {\sl Phys. Rev. Lett.} {\bf 77}, 2538 (1996) 

\bibitem{Grayson} 
  M.~Grayson, D.C.~Tsui, L.N.~Pfeiffer, K.W.~West, and A.M.~Chang, 
  {\sl Phys. Rev. Lett.} {\bf 80}, 1062 (1998) 

\bibitem{HLR}
  B.I.~Halperin, P.A.~Lee, N.~Read, {\sl Phys. Rev.} {\bf B47}, 7312 (1993)


\bibitem{Jain} J. K. Jain, Phys. Rev. Lett. {\bf 63}, 199 (1989).

\bibitem{CFedge}
  D.B.~Chklovskii, {\sl Phys. Rev.} {\bf B51}, 9895 (1995);
  L.~Brey, {\sl Phys. Rev.} {\bf B50}, 11861 (1994)

\bibitem{SLH} A.V.~Shytov, L.S.~Levitov, B.I.~Halperin,
  Phys. Rev. Lett., {\bf 80}, 141 (1998)

\bibitem{ContiVignale} S.~Conti and G.~Vignale, 
J. Phys. - Condens. Mat. {\bf 10}: (50) L779-L786 (1998)

\bibitem{HanThouless} J.H.~Han and D.J.~Thouless,
  Phys. Rev. B, {\bf 55}, 1926 (1997);\\
J.~H.~Han, Phys. Rev. B, {\bf 56}, 15806 (1997)

\bibitem{ZulickeMacDonald} U.~Z\"ulicke and A.H.~MacDonald,
Phys. Rev. {\bf B60}, p. 1837 (1999)

\bibitem{Pruisken} A.M.M.~Pruisken, B.~Skoric, M.A.~Baranov, 
Phys. Rev. {\bf B60}, p. 16838 (1999)

\bibitem{Cheianov} A.~Alekseev, V.~Cheianov, A.P.~Dmitriev, 
V.Yu.~Kachorovskii,
Pis'ma Zh. \'Eksp. Teor. Fiz {\bf 72} (6), 481-486 (2000) 
[JETP Lett. {\bf 72} (6), 333-336 (2000)]; 
cond-mat/9904076 

\bibitem{LopezFradkin} A.~Lopez and E.~Fradkin,
Phys.Rev.{\bf B59}, 15323 (1999)

\bibitem{LeeWen} D.-H. Lee, X.-G. Wen, 
Edge tunneling in fractional quantum Hall regime,  
cond-mat/9809160

\bibitem{Khveshchenko}
D.V. Khveshchenko, Solid State Commun. {\bf 111}: (9), 501-505 (1999); 
cond-mat/9806270 

\bibitem{Chang-plateau} 
A.M.~Chang, M.K.~Wu, C.C.~Chi, L.N.~Pfeiffer, 
K.W.~West, 
{\sl Phys. Rev. Lett.} {\bf 86}, 143 (2000)

\bibitem{LevitovShytov}
  L.S.~Levitov, A.V.~Shytov, 
Pis'ma Zh. \'Eksp. Teor. Fiz {\bf 60} (3), 200-205 (1997) 
[{\sl JETP Lett.} {\bf 66}, 214 (1997)];\\
       also, in:
        {\sl Correlated Fermions and Transport in Mesoscopic Systems},
edited by T. Martin, G. Montambaux, and J. Tran Thanh Van
        (Editions Frontieres, 1996), p. 513

\bibitem{OnsagerPrinciple} 
  L.D.~Landau and E.M.~Lifshits, 
Statistical Physics, Part I, Ch.~XII, \S~125, p.389 (Pergamon Press, 1982)

\bibitem{HePlatzmanHalperin} S. He, P.M. Platzman, B.I. Halperin,
{\sl Phys. Rev. Lett.} {\bf 71}, 777 (1993)

\bibitem{KimWen} 
  Y.B.~Kim, X.-G.~Wen, {\sl Phys. Rev.} {\bf B50}, 8078 (1994)

\bibitem{ShytovThesis} 
A.V.~Shytov, Ph. D. Thesis, 
L.D. Landau Institute for Theoretical Physics, Chernogolovka, 1999

\bibitem{fnKhvesh} In Ref.\cite{Khveshchenko}, Khveshchenko 
carried out an approximate
calculation in which he considered, simultaneously, effects of an unscreened 
Coulomb
interaction and a nonlocal CF conductivity, on the charge spreading action 
and the
$I-V$ curve for $eV/\hbar$, above the CF scattering frequency. We have not 
been able to
make a direct comparison of our results with his, however. 
We note that Khveshchenko 
does not include a term in the action which correspond to the logarithm of the
free-fermion Green's function, nor does he subtract a term corresponding to 
$S_{\rm free}$.


\bibitem{AAL} B.L.~Altshuler, A.G.~Aronov, P.A.~Lee, Phys.Rev.Lett. {\bf 44}, 
1288 (1980)

\bibitem{Volkov} 
  V.A.~Volkov and S.A.~Mikhailov, 
Zh. \'Eksp. Teor. Fiz {\bf 94} (8), 214-241 (1988)
[{\sl Sov. Phys. JETP} {\bf 67} 1639 (1988)]

\bibitem{Wiener-Hopf}
   B.~Noble, {\sl Methods based on the Wiener-Hopf technique}, 
        (Pergamon Press, London, 1958)

\bibitem{composite-edge} 
A.H.~MacDonald, Phys.Rev.Lett. {\bf 64}, 222 (1990);\\
M.D.~Johnson and A.H.~MacDonald, Phys.Rev.Lett. {\bf 67}, 2060 (1991)

\bibitem{effective-mass}
  B.L.~Altshuler, L.B.~Ioffe, and A.J.~Millis, 
         {\sl Phys. Rev.} {\bf B50}, 14049 (1994);\\
  Y.B.~Kim, A.~Furusaki, X.-G.~Wen and P.A.~Lee, 
         {\sl Phys. Rev.} {\bf B50}, 17917 (1994)










\end{thebibliography}
\end{document}